 \documentclass[final,1p]{elsarticle}

 \usepackage{graphicx}

\newtheorem{theorem}{Theorem}
\newtheorem{definition}{Definition}

\newtheorem{lemma}{Lemma}

\usepackage{color}

\definecolor{light-gray}{gray}{0.5}

\usepackage{amssymb,amsmath}

\input xy
\xyoption{all}

\begin{document}

\newcommand{\match}[1]{{{\color{light-gray}\blacktriangleleft}\!\!\!\!\!\!\lhd\!\!\!\!{\color{white}\blacktriangleleft}\!\!\!\!\!\lhd^{\!\!\!\!\!^{#1}\ }}}

\newcommand{\plug}[1]{\stackrel{#1}{\blacktriangleleft}}

\newcommand{\matchsubs}[1]{{{\color{light-gray}\blacktriangleleft}\:\!\!\!\!\!\!\lhd\:\!\!\!\!{\color{white}\blacktriangleleft}\!\!\!\!\!\lhd^{\!\!\!\!^{^{#1}}\,}}}

\newcommand{\plugsubs}[1]{\,\blacktriangleleft^{\!\!\!\!\!^{^{#1}}}}

\begin{frontmatter}



\title{\LARGE\textsc{Combinatorial Algebra for \\ second-quantized Quantum Theory}}


\author[ifj]{P. Blasiak}
\ead{pawel.blasiak@ifj.edu.pl}

\author[lipn]{G.H.E. Duchamp}
\ead{ghed@lipn-univ.paris13.fr}

\author[lptmc,open]{A.I. Solomon}
\ead{a.i.solomon@open.ac.uk}

\author[ifj]{A. Horzela}
\ead{andrzej.horzela@ifj.edu.pl}

\author[lptmc]{K.A. Penson}
\ead{penson@lptmc.jussieu.fr}

\address[ifj]{H. Niewodnicza\'nski Institute of
Nuclear Physics, Polish Academy of Sciences\\
ul.\ Eliasza-Radzikowskiego 152, PL 31342 Krak\'ow, Poland\vspace{0.2cm}}

\address[lipn]{Institut Galil\'ee -- Universit\'e Paris-Nord , LIPN, CNRS UMR 7030\\
99 Av.\ J.-B.\ Clement, F-93430 Villetaneuse, France\vspace{0.2cm}}

\address[lptmc]{Universit\'e Pierre et Marie Curie, LPTMC, CNRS UMR 7600\\
4 pl.\ Jussieu, F 75252 Paris Cedex 05, France\vspace{0.2cm}}

\address[open]{The Open University, Physics and Astronomy Department\\
Milton Keynes MK7 6AA, United Kingdom}

\begin{abstract}
We describe an algebra $\mathcal{G}$ of diagrams which faithfully gives a diagrammatic representation of  the structures of both the Heisenberg-Weyl algebra $\mathcal{H}$ -- the associative algebra of the creation and annihilation operators of quantum mechanics -- and $\mathcal{U}(\mathcal{L}_\mathcal{H})$, the enveloping algebra of the Heisenberg Lie algebra $\mathcal{L}_{\mathcal{H}}$.  We show explicitly how $\mathcal{G}$ may be endowed with the structure of a Hopf algebra, which is also mirrored in the structure of  $\mathcal{U}(\mathcal{L}_\mathcal{H})$. While both $\mathcal{H}$ and $\mathcal{U}(\mathcal{L}_\mathcal{H})$ are images of $\mathcal{G}$, the  algebra $\mathcal{G}$ has a richer structure and therefore embodies a finer combinatorial realization of the creation-annihilation system, of which it  provides a concrete model.
\end{abstract}


\end{frontmatter}





\section{Introduction}

One's comprehension of abstract mathematical concepts often  goes via concrete models. In many cases  convenient representations are obtained by using  combinatorial objects. Their advantage comes from simplicity based on intuitive notions of enumeration, composition and decomposition which allow for insightful interpretations, neat pictorial arguments and constructions \cite{FlajoletBook,BergeronBook,AignerBook}. This makes the combinatorial perspective particularly attractive for quantum physics,
due to the latter's  active pursuit of a better understanding of fundamental phenomena.
An example of such an attitude is given by  Feynman diagrams, which provide a graphical representation of quantum processes; these diagrams  became a  tool of choice in quantum field theory \cite{WeinbergBook,MattuckBook}. Recently, we have witnessed major progress in this area which has led to a rigorous combinatorial treatment of the renormalization procedure \cite{KreimerBook,KreimerFard} -- this breakthrough came with the recognition of  Hopf algebra structure in the perturbative expansions \cite{Kreimer1998,ConnesKreimer1998,Kreimer2003}.
There are many other examples in which combinatorial concepts play a crucial role,  ranging from  attempts to understand peculiar features of quantum formalism to a  novel approach to calculus, \textit{e.g.} see \cite{Spekkens,Baez,Coecke2010,CoeckeDuncan2009,LouckBook,Rasetti} for just a few recent developments in theses directions.
In the present paper we  consider some common algebraic structures of Quantum Theory and will show that the combinatorial approach has much to offer in this domain as well.

The current formalism and structure of Quantum Theory is based  on the theory of operators acting on a Hilbert space.
According to a few basic postulates, the physical concepts of a system, \textit{i.e.} the observables and transformations, find their representation as operators which account for experimental results.
An important role in this abstract description is played by  the notions of addition, multiplication and tensor product which are responsible for peculiarly quantum properties such as interference, non-compatibility of measurements as well as  entanglement in composite systems \cite{IshamBook,PeresBook,HughesBook}. From the algebraic point of view, one  appropriate structure capturing these features is a \emph{bi-algebra} or, more specifically,  a \textit{Hopf algebra}. These structures comprise a vector space with two operations, multiplication and co-multiplication, describing how operators compose and decompose. In the following, we shall be concerned with a combinatorial model which provides  an intuitive picture of this type of abstract structure.

However, the bare formalism is, by itself, not enough to provide  a description of real quantum phenomena. One must also associate operators with physical quantities.  This will, in turn, involve the association of some algebraic structure with physical concepts related to the system. In practice the most common correspondence rules are based on an associative algebra, the  \emph{Heisenberg-Weyl algebra} $\mathcal{H}$. This mainly arises by analogy with classical mechanics whose Poissonian structure is reflected in the quantum-mechanical commutator
of position and momentum observables $[x,p]=i\hbar$ \cite{DiracBook}.  In the first instance  this commutator  gives rise to a {\em Lie algebra} $\mathcal{L}_{\mathcal{H}}$ \cite{GilmoreBook,HallBook}, which naturally extends to a Hopf algebra structures in the enveloping algebra $\mathcal{U}(\mathcal{L}_{\mathcal{H}})$ \cite{BourbakiLieGroupsI,AbeBook}. An important equivalent commutator is that of the creation--annihilation operators  $[a,a^\dag]=1$, employed in the occupation number representation in quantum mechanics and the second quantization formalism of quantum field theory. Accordingly, we take the Heisenberg-Weyl  algebra  $\mathcal{H}$ as our starting point.

In this paper we  develop a combinatorial approach to  the Heisenberg-Weyl algebra
and present a comprehensive model of this algebra in terms of diagrams. In some respects this approach draws on  Feynman's idea of representing  physical processes as diagrams used as a bookkeeping tool in the perturbation expansions of quantum field theory. We discuss natural notions of diagram composition and decomposition which provide a straightforward interpretation of the abstract operations of multiplication and co-multiplication. The resulting  combinatorial algebra $\mathcal{G}$ may be seen as a lifting of the Heisenberg-Weyl algebra $\mathcal{H}$ to a richer structure of diagrams, capturing all the properties of the latter. Moreover, it will be shown to have a natural bi-algebra and Hopf algebra structure providing a concrete model for the enveloping algebra $\mathcal{U}(\mathcal{L}_\mathcal{H})$ as well. Schematically, these relationships can be pictured as follows
\begin{eqnarray}\nonumber
\xymatrix{
& \mathcal{G} \ar@{->>}[ld]_\varphi \ar@{->>}[rd]^{\bar{\varphi}}& &{\begin{array}{c}
Combinatorial\\ Algebra
\end{array}}\\
\mathcal{U}(\mathcal{L}_{\mathcal{H}}) \ar@{->>}[rr]^\pi && \mathcal{H}&Algebra\\
& \mathcal{L}_{\mathcal{H}} \ar@{^{(}->}[ru]_\kappa\ar@{_{(}->}[lu]^\iota&& Lie\ Algebra
}
\end{eqnarray}
where all the arrows are algebra morphisms and $\varphi$ is a Hopf algebra morphism. Whilst the lower part of the diagram is standard, the upper part and the construction of the combinatorial algebra $\mathcal{G}$ illustrate a genuine combinatorial underpinning of these abstract algebraic structures.

The paper is organized as follows. In Section \ref{HWalg} we start by briefly recalling the algebraic structure of the Heisenberg-Weyl algebra $\mathcal{H}$ and the enveloping algebra $\mathcal{U}(\mathcal{L}_{\mathcal{H}})$. In Section \ref{HWDiag} we define the Heisenberg-Weyl diagrams and introduce the notion of composition which leads to the combinatorial algebra $\mathcal{G}$. Section \ref{HWDiagDecomp} deals with the concept of decomposition, endowing the diagrams with a Hopf algebra structure. The relation between the combinatorial structures in $\mathcal{G}$ and the algebraic structures in $\mathcal{H}$ and $\mathcal{U}(\mathcal{L}_{\mathcal{H}})$ are explained as they appear in the construction. For ease of reading most  proofs have been  moved to the Appendices.

\section{Heisenberg-Weyl Algebra}\label{HWalg}

The objective of this paper is to develop a combinatorial model of the Heisenberg-Weyl algebra. In order to fully appreciate the versatility of our construction, we start by briefly recalling some common algebraic structures and clarifying their relation to the Heisenberg-Weyl algebra.

\subsection{Algebraic setting}

An \emph{associative algebra with unit} is one of the most basic structures used in the theoretical description of physical phenomena. It consists of a \emph{vector space} $\mathcal{A}$ over a field $\mathbb{K}$ which is equipped with a bilinear \emph{multiplication} law $\mathcal{A}\times\mathcal{A}\ni(x,y)\longrightarrow x\,y\in\mathcal{A}$
which is associative and possesses a \emph{unit} element $I$.\footnote{A full list of axioms may be found in any standard text on algebra, such as \cite{ArtinBook,BourbakiAlgebraI}.} Important notions in this framework are a \emph{basis} of an algebra, by which is meant  a basis for its underlying vector space structure, and the associated \emph{structure constants}. For each basis $\{x_i\}$ the latter are defined as the coefficients $\gamma_{ij}^k\in\mathbb{K}$ in the expansion of the product $x_i\,x_j=\sum_k \gamma_{ij}^k \,x_k$. We note that the structure constants uniquely determine the multiplication law in the algebra.\footnote{The structure constants must of course satisfy the constraints provided by the associative law.} For example, when the underlying vector space is finite dimensional of dimension $N$, that is each vector-space element  has a unique  expansion in terms of $N$ basis elements, then  there is only a finite number, at most $N^3$, of non-vanishing $\gamma_{ij}^k$'s.
A canonical example of the (noncommutative) associative algebra with unit is a matrix algebra, or more generally an algebra of linear operators acting in a vector space.

A description of composite systems is obtained through the construction of a tensor product. Of particular importance for physical applications is how the  transformations distribute among the components. A canonical example is the algebra of angular momentum and its representation on composite systems. In general, this issue is properly captured by the notion of a \emph{bi-algebra} which consists of an associative algebra with unit $\mathcal{A}$ which is additionally equipped with a \emph{co-product} and a \emph{co-unit}. The co-product is defined as a co-associative
linear mapping $\Delta:\mathcal{A}\longrightarrow\mathcal{A}\otimes\mathcal{A}$ prescribing the action from the  algebra to a tensor product, whilst the co-unit $\varepsilon:\mathcal{A}\longrightarrow\mathbb{K}$ gives a linear map to the  underlying field $\mathbb{K}$. Furthermore, the bi-algebra axioms require $\Delta$ and $\varepsilon$ to be algebra morphisms, \textit{i.e.} to preserve multiplication in the algebra, which asserts the correct transfer of the algebraic structure of $\mathcal{A}$ into the tensor product $\mathcal{A}\otimes\mathcal{A}$. Additionally, a proper description of the action of an algebra in a dual space requires the existence of an antimorphism $S:\mathcal{A}\longrightarrow\mathcal{A}$ called the \emph{antipode}, thus  introducing a \emph{Hopf algebra} structure in $\mathcal{A}$. For a complete set of bi-algebra and Hopf algebra axioms see \cite{SweedlerBook,AbeBook,CartierHopfPrimer}.

In this context it is instructive  to discuss the difference between Lie algebras and associative algebras which is often misunderstood.
A \emph{Lie algebra} is a vector space $\mathcal{L}$ over a field $\mathbb{K}$ with a bilinear law $\mathcal{L}\times\mathcal{L}\ni(x,y)\longrightarrow[x,y]\in\mathcal{L}$, called the Lie bracket, which is antisymmetric $[x,y]=-[y,x]$ and satisfies the Jacobi identity: $\left[x,[y,z]\right]+\left[y,[z,x]\right]+\left[z,[x,y]\right]=0$. Lie algebras are not associative in general\footnote{However, all the Heisenberg Lie algebras $h_{2n+1}$ are also (trivially) associative in the sense that for all $x,y,z \in h_{2n+1}$, $x\star(y\star z)= (x \star y) \star z\,(=0)$ where $\star$ is the composition (bracket) in the Lie algebra.} and lack an identity element. A standard remedy for these deficiencies consists of passing to the \emph{enveloping algebra} $\mathcal{U}(\mathcal{L})$ which has the more familiar structure of an associative algebra with unit and, at the same time, captures all the relevant properties of $\mathcal{L}$. An important step in its realization is the Poincar\'e-Birkhoff-Witt theorem which provides an explicit description of $\mathcal{U}(\mathcal{L})$ in terms of ordered monomials in the basis elements of $\mathcal{L}$, see \cite{BourbakiLieGroupsI}. As such, the enveloping algebras can be seen as giving faithful models of Lie algebras in terms of a structure with an associative law.

Below, we  illustrate these abstract algebraic constructions within the context of the Heisenberg-Weyl algebra. These abstract algebraic concepts gain by use of  a concrete example.

\subsection{Heisenberg-Weyl algebra revisited}

In this paper we consider the \emph{Heisenberg-Weyl algebra}, denoted by $\mathcal{H}$, which is an associative algebra with unit, generated by two elements $a$ and $a^\dag$ subject to the relation
\begin{eqnarray}\label{aa}
a\,a^\dag=a^\dag a+I\ .
\end{eqnarray}
This means that the algebra consists of elements $A\in\mathcal{H}$ which are linear combinations of finite products of the generators, \textit{i.e.}
\begin{eqnarray}\label{Aaaaaa}
A=\sum_{\substack{r_k,...,r_1\\s_k,...,s_1}}A_{\substack{r_k,...,r_1\\s_k,...,s_1}}\ \ a^{\dag\,r_k}\, a^{s_k}\,...\ a^{\dag\,r_2}\, a^{s_2}\ a^{\dag\,r_1}\, a^{s_1},
\end{eqnarray}
where the sum ranges over a finite set of multi-indices $r_k,...,r_1\in\mathbb{N}$ and $s_k,...,s_1\in\mathbb{N}$ (with the convention $a^0=a^{\dag\,0}=I$).
Throughout the paper we stick to the notation used in the occupation number representation in which $a$ and $a^\dag$ are interpreted as \emph{annihilation} and \emph{creation} operators. We note, however, that one should not attach too much weight to this choice as we consider algebraic properties only, so particular realizations are irrelevant and the crux of the study is the sole relation of Eq.~(\ref{aa}). For example, one could equally well use $X$ as  multiplication by $z$, and derivative operator $D=\partial_z$  acting in the space of complex polynomials, or analytic functions, which also satisfy the relation $[D,X]=I$.

Observe that the representation given by Eq.~(\ref{Aaaaaa}) is ambiguous in so far as  the rewrite rule of Eq.~(\ref{aa}) allows different representations of the same element of the algebra, \textit{e.g.} $aa^\dag$ or equally $a^\dag a +I$.
The remedy for this situation lies in fixing a preferred order of the generators.
Conventionally, this is done by choosing the \emph{normally ordered} form in which
all annihilators stand to the right of creators.
As a result, each element of the algebra $\mathcal{H}$ can be uniquely written in  normally ordered form as
\begin{eqnarray}\label{A}
A=\sum_{k,l}\alpha_{kl}\ a^{\dag\,k}\, a^l\,.
\end{eqnarray}
In this way, we find  that the normally ordered monomials constitute a natural basis for the Heisenberg-Weyl algebra, \emph{i.e.}
\begin{eqnarray}\nonumber
\textsc{Basis\ of\ } \mathcal{H}\,: \ \ \ \ \ \ \left\{\,a^{\dag\,k} a^l\,\right\}_{k,l\in\mathbb{N}}\,,
\end{eqnarray}
indexed by pairs of integers $k,l=0,1,2,...$, and Eq.~(\ref{A}) is the expansion of the element $A$ in this basis. One should note that the normally ordered representation of the elements of the algebra suggests itself not only as the simplest one but is also of practical use and importance in applications in quantum optics \cite{Glauber,Schleich,KlauderBook} and quantum field theory \cite{BjorkenDrell,MattuckBook}. In the sequel we choose to work in this particular basis. For the complete algebraic description of $\mathcal{H}$ we still need the structure constants of the algebra. They can be readily read off from the formula for the expansion of the product of basis elements
\begin{eqnarray}\label{HWproduct}
a^{\dag\,p} a^q \, a^{\dag\,k} a^l  =
\sum_{i=0}^{\text{min}\,\{q,k\}} \binom{q}{i}\binom{k}{i}\, i!\ a^{\dag\,p+k-i} a^{q+l-i}\ .
\end{eqnarray}
We note that working in a fixed basis is in general a nontrivial task. In our case, the problem reduces to rearranging $a$ and $a^\dag$ to  normally ordered form which may often  be achieved  by  combinatorial methods \cite{Wilcox,AmJPhys}.

\subsection{Enveloping algebra $\mathcal{U}(\mathcal{L}_\mathcal{H})$}

We recall that the \emph{Heisenberg Lie algebra}, denoted by $\mathcal{L}_\mathcal{H}$,\footnote{This  Lie algebra, the Heisenberg Lie algebra, which is written here as  $\mathcal{L}_{\mathcal{H}}$ ,  is often called $h_3$ in the literature, with $h_{2n+1}$ being the extension to $n$ creation operators.} is a 3-dimensional vector space with basis $\{a^\dag,a,e\}$ and  Lie bracket defined by $[a,a^\dag]=e$, $[a^\dag,e]=[a,e]=0$. Passing to the enveloping algebra involves imposing the linear order $a^\dag\succ a \succ e$ and constructing the \emph{enveloping algebra} $\mathcal{U}(\mathcal{L}_\mathcal{H})$ with  basis given by the family
\begin{eqnarray}\nonumber
\textsc{Basis\ of\ } \mathcal{U}(\mathcal{L}_\mathcal{H})\,: \ \ \ \ \ \ \left\{\,a^{\dag\,k} a^l\, e^m\,\right\}_{k,l,m\in\mathbb{N}}\,,
\end{eqnarray}
which is indexed by triples of integers $k,l,m=0,1,2,...$.
Hence, elements $B\in\mathcal{U}(\mathcal{L}_\mathcal{H})$ are of the form
\begin{eqnarray}\label{B}
B=\sum_{k,l,m}\beta_{klm}\ a^{\dag\,k} a^l\,e^m\,.
\end{eqnarray}
According to the Poincar\'e-Birkhoff-Witt theorem, the associative multiplication law in the enveloping algebra $\mathcal{U}(\mathcal{L}_\mathcal{H})$ is defined by concatenation,  subject to the rewrite rules
\begin{eqnarray}\nonumber
a\, a^\dag&=&a^\dag a+e\, ,\\\label{Uaa}
e\, a^\dag&=&a^\dag e\, ,\\\nonumber
e\, a\ &=&a\,e\ .
\end{eqnarray}
One checks that the formula for multiplication of basis elements in $\mathcal{U}(\mathcal{L}_\mathcal{H})$ is a slight generalization of Eq.~(\ref{HWproduct}) and is
\begin{eqnarray}\label{StrConst}
a^{\dag\,p} a^q\, e^r\, a^{\dag\,k} a^l\, e^m =
\sum_{i=0}^{\text{min}\,\{q,k\}} \binom{q}{i}\binom{k}{i}\, i!\ a^{\dag\,p+k-i}\, a^{q+l-i}\, e^{r+m+i}\ .
\end{eqnarray}

Note that the algebra $\mathcal{U}(\mathcal{L}_\mathcal{H})$ differs from $\mathcal{H}$ by the additional central element $e$ which should not be confused with the unity $I$ of the enveloping algebra.\footnote{As usual, we write $a^0={a^{\dag}}^0=e^0=I$} This distinction plays an important role in some applications as explained below. In situations when this difference is insubstantial one may set $e\rightarrow I$ recovering the Heisenberg-Weyl algebra $\mathcal{H}$, \textit{i.e.} we have the surjective morphism $\pi:\mathcal{U}(\mathcal{L}_\mathcal{H})\longrightarrow\mathcal{H}$ given by
\begin{eqnarray}
\pi\left(a^{\dag\,i} a^j\, e^k\right)=a^{\dag\,i} a^j .
\end{eqnarray}
This completes the algebraic picture which can be subsumed in the following diagram
\begin{eqnarray}\nonumber
\xymatrix{
\mathcal{U}(\mathcal{L}_{\mathcal{H}}) \ar@{->>}[rr] ^\pi&& \mathcal{H}\\
& \mathcal{L}_{\mathcal{H}} \ar@{^{(}->}[ru]_\kappa\ar@{_{(}->}[lu]^\iota&&
}
\end{eqnarray}
We emphasize that the inclusions $\iota:\mathcal{L}_\mathcal{H}\longrightarrow\mathcal{U}(\mathcal{L}_\mathcal{H})$ and  $\kappa=\pi\circ\iota:\mathcal{L}_\mathcal{H}\longrightarrow\mathcal{H}$ are Lie algebra morphisms, while the surjection $\pi:\mathcal{U}(\mathcal{L}_\mathcal{H}) \longrightarrow\mathcal{H}$ is a morphism of associative algebras with unit. Note that different structures are carried over by these morphisms.

Finally, we observe that the enveloping algebra $\mathcal{U}(\mathcal{L}_{\mathcal{H}})$ may be equipped with a Hopf algebra structure. This may be  constructed in a standard way by defining the co-product\footnote{Note that this definition gives a co-commutative Hopf algebra. One may also define a non-co-commutative co-product \cite{Ballesteros}.}
 $\Delta:\mathcal{U}(\mathcal{L}_{\mathcal{H}})\longrightarrow\mathcal{U}(\mathcal{L}_{\mathcal{H}})\otimes\mathcal{U}(\mathcal{L}_{\mathcal{H}})$ on the generators $x=a^\dag,\,a,\,e$ setting $\Delta(x)=x\otimes I+I\otimes x$, which further extends to
\begin{eqnarray}\label{Uco-product}
\Delta\left(a^{\dag\,p} a^q\, e^r\right)
=\sum_{i,j,k}\binom{p}{i}\binom{q}{j}\binom{r}{k}\ a^{\dag\,i} a^j\, e^k\otimes a^{\dag\,{p-i}} a^{q-j}\, e^{r-k}\,.
\end{eqnarray}
Similarly, the antipode $S:\mathcal{U}(\mathcal{L}_{H})\longrightarrow\mathcal{U}(\mathcal{L}_{H})$ is given on generators by $S(x)=-x$, and hence from the anti-morphism property yields
\begin{eqnarray}\label{Uantipode}
S\left(a^{\dag\,p} a^q\, e^r\right)=(-1)^{p+q+r}\ e^r\, a^q\, a^{\dag\,p}\,.
\end{eqnarray}
Finally, the co-unit $\varepsilon:\mathcal{U}(\mathcal{L}_{H})\longrightarrow\mathbb{K}$ is defined in the following way
\begin{eqnarray}\label{Uco-unit}
\varepsilon\left(a^{\dag\,p} a^q\, e^r\right)&=&
\left\{
\begin{array}{l}
1 \text{\ \ \ \ if\ \ \ }p,q,r=0\ ,\\
0  \text{\ \ \ \ otherwise\ . }
\end{array}\right.
\end{eqnarray}
A word of warning here: the Heisenberg-Weyl algebra $\mathcal{H}$ can not be endowed with a bi-algebra structure contrary to what is sometimes tacitly assumed. This  is because properties of the co-unit contradict the relation of Eq.~(\ref{aa}), \textit{i.e.} it follows that $\varepsilon(I)=\varepsilon(a\,a^\dag-a^\dag a)=\varepsilon(a)\,\varepsilon(a^\dag)-\varepsilon(a^\dag)\,\varepsilon(a)=0$ whilst one should have $\varepsilon(I)=1$.
This brings out the importance of the additional central element $e\neq I$ which saves the day for $\mathcal{U}(\mathcal{L}_\mathcal{H})$.

\section{Algebra of Diagrams and composition}\label{HWDiag}

In this Section we define the combinatorial class of Heisenberg-Weyl diagrams which is the central object of our study. We equip this class  with an intuitive notion of {\em composition}, permitting the  construction of an algebra structure and thus  providing a combinatorial model of the algebras $\mathcal{H}$ and $\mathcal{U}(\mathcal{L}_{\mathcal{H}})$.

\subsection{Combinatorial concepts}

We start by recalling a few basic notions from graph theory \cite{DiestelBook} needed for a precise definition of the Heisenberg-Weyl diagrams, and then provide an intuitive graphical representation of this structure.

Briefly, from a set-theoretical point of view, a \emph{directed graph} is a collection of  \emph{edges} $E$ and \emph{vertices} $V$ with the structure determined by two mappings $h,t:E\longrightarrow V$ prescribing how the \emph{head} and \emph{tail} of an edge are attached to vertices. Here we address a slightly more general setting consisting of \emph{partially defined} graphs whose edges may have one of the ends free (but not both), \textit{i.e.} we  consider finite graphs with partially defined mappings $h$ and $t$ such that $dom(h)\cup dom(t)=E$, where $dom$ stands for domain. We call a \emph{cycle} in a graph any sequence of edges $e_1,e_2,...,e_n$ such that $h(e_k)=t(e_{k+1})$ for $k<n$ and $h(e_n)=t(e_1)$. A convenient concept in graph theory concerns the notion of equivalence. Two graphs given by $h_1,t_1:E_1\longrightarrow V_1$ and $h_2,t_2:E_2\longrightarrow V_2$ are said to be \emph{equivalent} if one can be isomorphically transformed into the other, \textit{i.e.} both have the same number of vertices and edges and there exist two isomorphisms $\alpha_E:E_1\longrightarrow E_2$ and $\alpha_V:V_1\longrightarrow V_2$ faithfully transferring the structure of the graphs in the following sense
\begin{eqnarray}\nonumber
\xymatrix{
E_1\ \ar@{->}@< 2pt> [rr] ^h \ar@{->}@<-2pt> [rr]_t \ar@{->}[d]_{\alpha_E}&&\ V_1\ar@{->}[d]^{\alpha_V}\\
E_2\ \ar@{->}@< 2pt> [rr] ^h \ar@{->}@<-2pt> [rr]_t&&\ V_2
}
\end{eqnarray}
The advantage of  \emph{equivalence classes} so defined  is that we can liberate ourselves from specific set-theoretical realizations and think of a graph only in terms of relations between vertices and edges which can be conveniently described in a graphical way -- this is the attitude we adopt in the sequel.

In this context, we propose the following formal definition:

\begin{definition}[Heisenberg-Weyl Diagrams] \label{HWDiag-def} \ \vspace{0.1cm}\\
A Heisenberg-Weyl diagram $\varGamma$ is a class of partially defined directed graphs without cycles. It consists of three sorts of lines: the inner ones $\varGamma^{\stackrel{0}{}}$ having both head and tail attached to vertices, the incoming lines $\varGamma^{\stackrel{-}{}}$ with free tails, and the outgoing lines $\varGamma^{\stackrel{+}{}}$ with free heads.
\end{definition}

A typical {\it modus operandi} when working with classes is to invoke representatives. Following this practice, by default we make all statements concerning Heisenberg-Weyl diagrams with reference to  its representatives, assuming that they are class invariants,  which assumption can be routinely checked in each case.

\begin{figure}[h]
\begin{center}
\resizebox{0.6\textwidth}{!}{\includegraphics{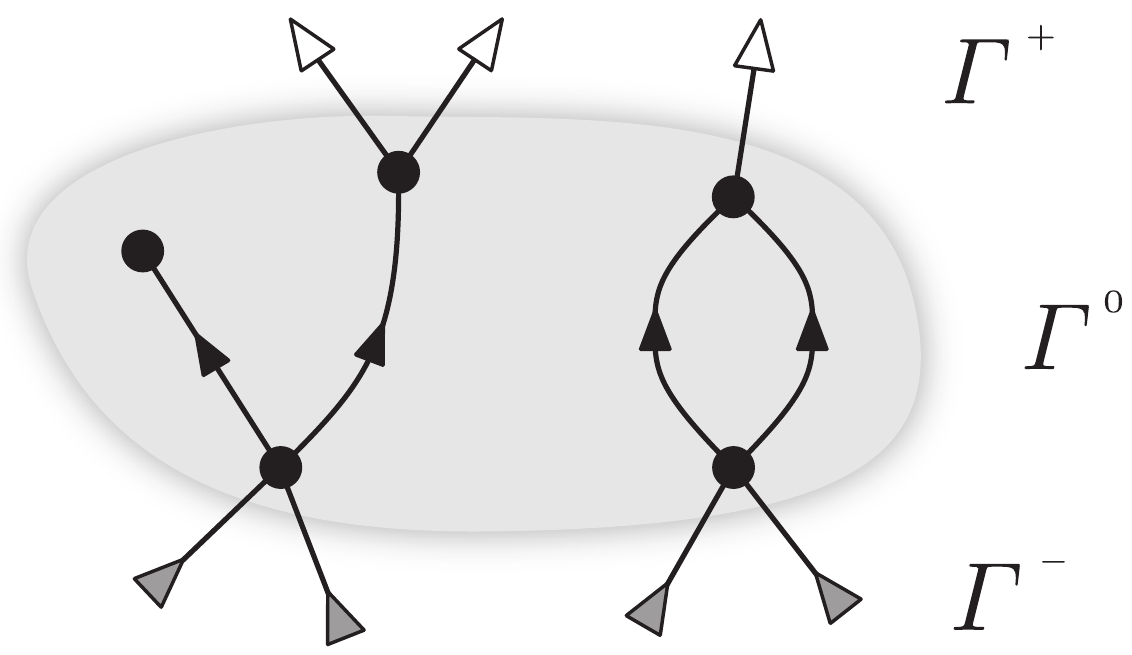}}
\caption{\label{Diagram-Def} An example of a Heisenberg-Weyl diagram with three distinguished  characteristic sorts of lines: the inner ones $|\varGamma^{\stackrel{0}{}}|=4$, the incoming lines $|\varGamma^{\stackrel{-}{}}|=4$ and outgoing lines $|\varGamma^{\stackrel{+}{}}|=3$.}
\end{center}
\end{figure}

The formal Definition~\ref{HWDiag-def} gives an intuitive picture in graphical form -  see the illustration Fig.~\ref{Diagram-Def}. A diagram can be represented as a set of vertices $\bullet$ connected by lines each carrying an arrow indicating the direction from the tail to the head. Lines having one of the ends not attached to a vertex will be marked with $\vartriangle$ or ${{\color{light-gray}\blacktriangle}\!\!\!\!\!\!\vartriangle}$ at the free head or tail respectively.
We will conventionally draw all incoming lines at the bottom and the outgoing lines at the top with all arrows heading upwards; this is always possible since the diagrams do not have cycles. This pictures the Heisenberg-Weyl diagram as a sort of process or transformation with vertices playing the role of intermediate steps.
\bigskip

An important characteristic of a diagram $\Gamma$ is the total number of its lines denoted by $|\varGamma|$. In the next sections we further refine counting of the lines to the inner, the incoming and the outgoing lines, denoting the result by $|\varGamma^{\stackrel{0}{}}|$, $|\varGamma^{\stackrel{-}{}}|$ and $|\varGamma^{\stackrel{+}{}}|$ respectively. Clearly, one has $|\varGamma|=|\varGamma^{\stackrel{0}{}}|+|\varGamma^{\stackrel{-}{}}|+|\varGamma^{\stackrel{+}{}}|$.

\subsection{Diagram composition}\label{DiagramComposition}

A crucial concept of this paper concerns {\em composition} of Heisenberg-Weyl diagrams. This has  a straightforward graphical representation as the attaching of free lines one to another, and is based on the notion of a matching.

A \emph{matching} $m$ of two sets $A$ and $B$ is a choice of pairs $(a_i,b_i)\in A\times B$ all having different components, \textit{i.e.} if $a_i=a_j$ or $b_i=b_j$ then $i=j$. Intuitively, it is a collection of pairs $(a_i , b_i )$ obtained by taking away $a_i$ from $A$ and $b_i$ from $B$ and repeating the process several times with sets $A$ and $B$ gradually reducing in size. We denote the collection of all possible matchings by $A\match{\ } B$, and its restriction to matchings comprising $i$ pairs only by $A\match{i}B$. It is straightforward to check by exact enumeration the formula $|A\match{i}B|=\binom{|A|}{i}\binom{|B|}{i}\ i!$,
which is valid for any $i$ if the convention $\binom{n}{k}=0$ for $n<k$ is applied.

The concept of diagram composition suggests itself, as:

\begin{definition}[Diagram Composition] \label{HWDiag-comp}\ \vspace{0.15cm}\\
Consider two Heisenberg-Weyl diagrams $\varGamma_2$ and $\varGamma_1$ and a matching $m\in\varGamma^{\stackrel{-}{}}_2\match{\ }\varGamma^{\stackrel{+}{}}_1$ between the free lines going out from the first one $\varGamma^{\stackrel{+}{}}_1$ and the free lines going into the second one $\varGamma^{\stackrel{-}{}}_2$. The composite diagram, denoted by $\varGamma_2\plug{m}\varGamma_1$, is constructed by joining the lines coupled by the matching $m$.
\end{definition}

\begin{figure}[h]
\begin{center}
\resizebox{0.4\textwidth}{!}{\includegraphics{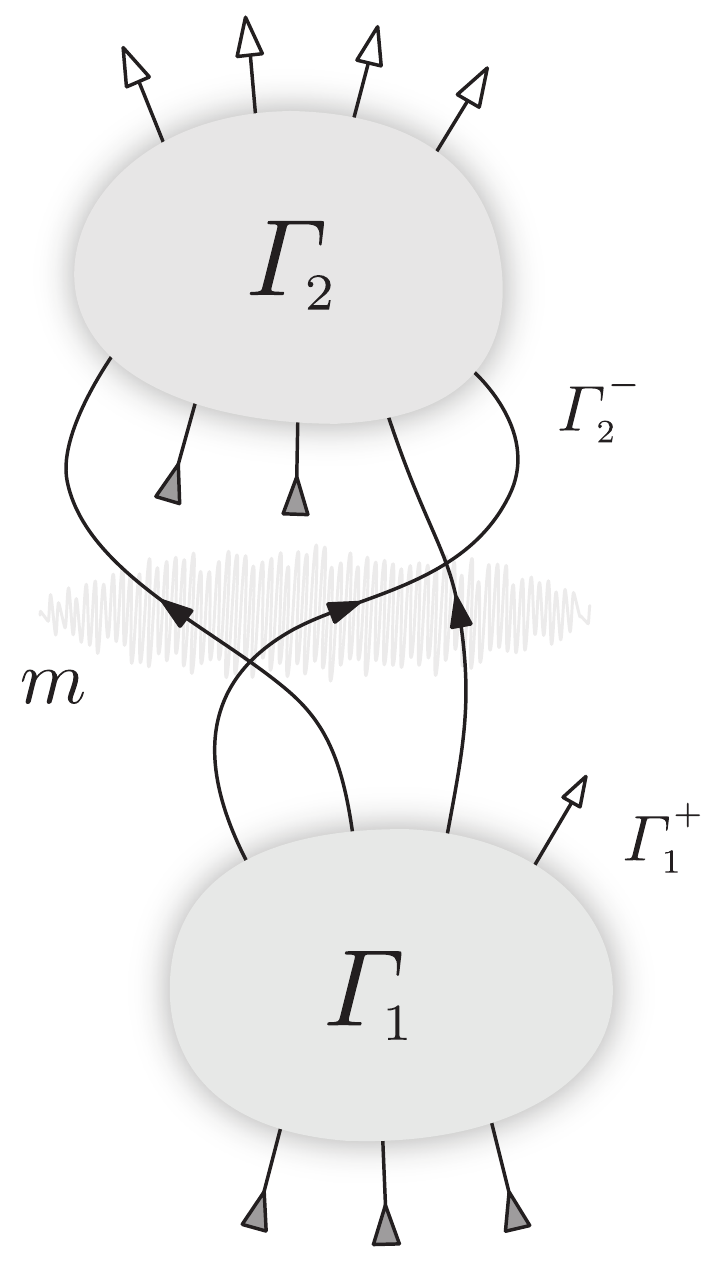}}
\caption{\label{Diagram-comp} Composition of two diagrams $\varGamma_2\plug{m}\varGamma_1$ according to the matching $m\in\varGamma^{\stackrel{-}{}}_2\match{\ }\varGamma^{\stackrel{+}{}}_1$ consisting of three connections.}
\end{center}
\end{figure}

This descriptive definition can be formalized by referring to representatives in the following way.  Given two disjoint graphs $\varGamma_1$ and $\varGamma_2$, \textit{i.e.} such that $V_{\varGamma_2}\cap V_{\varGamma_1}=\text{\O}$ and $E_{\varGamma_2}\cap E_{\varGamma_1}=\text{\O}$, we construct the composite graph $\varGamma_2\plug{m}\varGamma_1$ consisting of vertices $V_{\varGamma_2\plugsubs{m}\varGamma_1}=V_{\varGamma_2}\cup V_{\varGamma_1}$ and edges $E_{\varGamma_2\plugsubs{m}\varGamma_1}=E_{\varGamma_2}\cup E_{\varGamma_1}\cup m - \left(pr_{2}(m)\cup pr_{1}(m)\right)$, where $pr$ is the projection on the first or second component in  $E_{\varGamma_2}\times E_{\varGamma_1}$. Then, the head and tail functions unambiguously extend to the set $E_{\varGamma_2}\cup E_{\varGamma_1}- \left(pr_{2}(m)\cup pr_{1}(m)\right)$ and for $e=(e_{\varGamma_2},e_{\varGamma_1})\in m$ we define $h_{\varGamma_2\plugsubs{m}\varGamma_1}(e)=h_{\varGamma_2}(e_{\varGamma_2})$ and $t_{\varGamma_2\plugsubs{m}\varGamma_1}(e)=t_{\varGamma_1}(e_{\varGamma_1})$. Clearly, choice of the disjoint graphs in classes is always possible and the resulting directed graph does not contain cycles. It then  remains to check that  the composition of diagrams so defined, making use of representatives, is class invariant.

Definition~\ref{HWDiag-comp} can be straightforwardly seen as if diagrams were put  over one another with some of the lines going out from the lower one plugged into some of the lines going into  the upper one in accordance with a given matching $m\in\varGamma^{\stackrel{-}{}}_2\match{\ }\varGamma^{\stackrel{+}{}}_1$, for illustration see Fig.~\ref{Diagram-comp}. Observe that in general two graphs can be composed in many ways, \textit{i.e.} as many as there are possible matchings (elements in $\varGamma^{\stackrel{-}{}}_2\match{\ }\varGamma^{\stackrel{+}{}}_1$). In Section~\ref{AlgebraHWDiag} we exploit all these possible compositions to endow the diagrams with the structure of an algebra. Note also that the above construction depends on the order in which diagrams are composed and in general the reverse order yields different results.

We conclude by two simple remarks concerning the composition of two diagrams $\varGamma_2$ and $\varGamma_1$ constructed by joining exactly $i$ lines.  Firstly, we observe that possible compositions can be enumerated explicitly by the formula
\begin{eqnarray}\label{num-diag-match}
|\varGamma^{\stackrel{-}{}}_2\match{i}\varGamma^{\stackrel{+}{}}_1|=\binom{|\varGamma_2^{\stackrel{-}{}}|}{i}\binom{|\varGamma_1^{\stackrel{+}{}}|}{i}\ i!\ .
\end{eqnarray}
Secondly, the number of incoming, outgoing and inner lines in the composed diagram does not depend on the choice of a matching $m\in\varGamma^{\stackrel{-}{}}_2\match{i}\varGamma^{\stackrel{+}{}}_1$ and reads respectively
\begin{eqnarray}\label{phi-of-match}
|(\varGamma_2\plug{m} \varGamma_1)^{\stackrel{+}{}}|&=&|\varGamma_2^{\stackrel{+}{}}|+|\varGamma_1^{\stackrel{+}{}}|-i\ ,\nonumber\\
|(\varGamma_2\plug{m} \varGamma_1)^{\stackrel{-}{}}|&=&|\varGamma_2^{\stackrel{-}{}}|+|\varGamma_1^{\stackrel{-}{}}|-i\ ,\nonumber\\
|(\varGamma_2\plug{m} \varGamma_1)^{\stackrel{0}{}}\,|&=&|\varGamma_2^{\stackrel{0}{}}\,|+|\varGamma_1^{\stackrel{0}{}}\,|+i\ .
\end{eqnarray}

\subsection{Algebra of Heisenberg-Weyl Diagrams}\label{AlgebraHWDiag}

We show here that the Heisenberg-Weyl diagrams come equipped
with a natural algebraic structure based on diagram composition. It will appear to be a combinatorial refinement of the familiar algebras $\mathcal{H}$ and $\mathcal{U}(\mathcal{L}_{\mathcal{H}})$.

An algebra requires two operations, addition and multiplication, which we construct in the following way.
We define $\mathcal{G}$ as a vector space over $\mathbb{K}$ generated by the basis set consisting of all Heisenberg-Weyl diagrams, \textit{i.e.}
\begin{eqnarray}\label{G}
\mathcal{G}=\left\{\ {\sum}_i\alpha_i\ \varGamma_i:\ \alpha_i\in\mathbb{K},\ \varGamma_i-\text{Heisenberg-Weyl\ diagram}\ \right\}.
\end{eqnarray}
Addition and multiplication by scalars in $\mathcal{G}$ has the usual form
\begin{eqnarray}\label{Addition}
{\sum}_i\ \alpha_i\ \varGamma_i+{\sum}_i\ \beta_i\ \varGamma_i={\sum}_i\ (\alpha_i+\beta_i)\ \varGamma_i\ ,
\end{eqnarray}
and
\begin{eqnarray}\label{MultiplicationScalars}
\beta\ {\sum}_i\ \alpha_i\ \varGamma_i={\sum}_i\ \beta\,\alpha_i\ \varGamma_i\ .
\end{eqnarray}
The nontrivial part in the definition of the algebra $\mathcal{G}$ concerns multiplication, which by bilinearity
\begin{eqnarray}
{\sum}_i\ \alpha_i\ \varGamma_i*{\sum}_j\ \beta_j\ \varGamma_j={\sum}_{i,j}\ \alpha_i\beta_j\ \varGamma_i*\varGamma_j,
\end{eqnarray}
reduces to determining it on the basis set of the Heisenberg-Weyl diagrams.
Recalling the notions of Section~\ref{DiagramComposition}, we define the product of two diagrams $\varGamma_2$ and $\varGamma_1$ as the sum of all possible compositions, \textit{i.e.}
\begin{eqnarray}\label{multiplication-def}
\varGamma_2*\varGamma_1=\sum_{m\in\varGamma^{\stackrel{-}{}}_2\matchsubs{\ }\varGamma^{\stackrel{+}{}}_1}\varGamma_2\plug{m}\varGamma_1\ .
\end{eqnarray}
Clearly, the sum is well defined as there is only a finite number of compositions (elements in $\varGamma^{\stackrel{-}{}}_2\match{\ }\varGamma^{\stackrel{+}{}}_1$). Note that although all coefficients in Eq.~(\ref{multiplication-def}) are equal to one, some terms in the sum may appear several times giving rise to nontrivial structure constants.
The multiplication thus  defined is noncommutative and possesses a unit element which is the empty graph \O\ (no vertices, no lines). Moreover, the following theorem holds (for the proof of associativity see Appendix~\ref{Appendix-Associativity}):
\begin{theorem}[Algebra of Diagrams] \label{Diag-AAU}\ \vspace{0.15cm}\\
Heisenberg-Weyl diagrams form a (noncommutative) associative algebra with unit $(\mathcal{G},+,*,\text{\emph{\O}})$.
\end{theorem}

Our objective, now, is to clarify the relation of the algebra of Heisenberg-Weyl diagrams $\mathcal{G}$ to the physically relevant algebras $\mathcal{U}(\mathcal{L}_{\mathcal{H}})$ and $\mathcal{H}$. We shall construct \emph{forgetful} mappings which give a simple combinatorial prescription of how to obtain the two latter structures from $\mathcal{G}$.

We define a linear mapping $\varphi:\mathcal{G}\longrightarrow\mathcal{U}(\mathcal{L}_{H})$ on the basis elements by
\begin{eqnarray}\label{phi}
\varphi(\varGamma)=a^{\dag\,{|\varGamma^{\,\stackrel{+}{}}\!|}}\ a^{\,|\varGamma^{\stackrel{-}{}}\!|}\ e^{\,|\varGamma^{\stackrel{0}{}}|}\ .
\end{eqnarray}
This prescription can be intuitively understood by looking at the diagrams as if they were carrying auxiliary labels $a^\dag$, $a$ and $e$ attached to all the outgoing, incoming and inner lines respectively. Then the mapping of Eq.~(\ref{phi}) just neglects the structure of the graph and only pays attention to the number of lines, \emph{i.e}. counting them according to the labels. Clearly, $\varphi$ is onto and it can be proved to be a genuine algebra morphism, \textit{i.e.} it preserves addition and multiplication in $\mathcal{G}$ (for the proof see Appendix~\ref{Appendix-Morphism}).

Similarly, we define the morphism $\bar{\varphi}:\mathcal{G}\longrightarrow \mathcal{H}$ as
\begin{eqnarray}\label{phi-bar}
\bar{\varphi}(\varGamma)=a^{\,\dag{|\varGamma^{\,\stackrel{+}{}}\!|}}\ a^{\,|\varGamma^{\stackrel{-}{}}\!|}\ ,
\end{eqnarray}
which differs from $\varphi$ by ignoring all inner lines in the diagrams. It can be expressed as $\bar{\varphi}=\pi\circ\varphi$ and hence satisfies  all the properties of an algebra morphism.

We recapitulate the above discussion in the following theorem:
\begin{theorem}[Forgetful mapping] \label{Forgetful-mapping}\ \vspace{0.1cm}\\
The mappings $\varphi:\mathcal{G}\longrightarrow\mathcal{U}(\mathcal{L}_{\mathcal{H}})$ and $\bar{\varphi}:\mathcal{G}\longrightarrow \mathcal{H}$ defined in Eqs.~(\ref{phi}) and (\ref{phi-bar}) are surjective algebra morphisms, and the following diagram commutes
\begin{eqnarray}
\xymatrix{
& \mathcal{G} \ar@{->>}[ld]_\varphi \ar@{->>}[rd]^{\bar{\varphi}}& \\
\mathcal{U}(\mathcal{L}_{\mathcal{H}}) \ar@{->>}[rr]^\pi && \mathcal{H}}
\end{eqnarray}
\end{theorem}
Therefore, the algebra of Heisenberg-Weyl diagrams $\mathcal{G}$ is a lifting of the algebras $\mathcal{U}(\mathcal{L}_{\mathcal{H}})$ and $\mathcal{H}$, and the latter two can be recovered by applying appropriate forgetful mappings $\varphi$ and $\bar{\varphi}$. As such, the algebra $\mathcal{G}$ can be seen as a fine graining of the abstract algebras $\mathcal{U}(\mathcal{L}_{\mathcal{H}})$ and $\mathcal{H}$. Thus these latter algebras  gain a concrete combinatorial interpretation in terms of the richer structure of diagrams.

\section{Diagram Decomposition and Hopf algebra}\label{HWDiagDecomp}

We have seen in Section~\ref{HWDiag} how the notion of composition allows for a combinatorial definition of diagram multiplication, opening the door to the realm of algebra. Here, we consider the opposite concept of diagram {\em decomposition} which induces a combinatorial co-product in the algebra, thus  endowing Heisenberg-Weyl diagrams with a bi-algebra structure. Furthermore, we will show that $\mathcal{G}$ forms a Hopf algebra as well.

\subsection{Basic concepts: Combinatorial decomposition}\label{BasicNotions}

Suppose we are  given a class of objects which allow for decomposition, \textit{i.e.}  split into ordered pairs of pieces from the same class. Without loss of generality one may think of the class of Heisenberg-Weyl diagrams and some, for the moment unspecified, procedure assigning to a given diagram $\varGamma$ its possible decompositions $(\varGamma'',\varGamma')$. In general there might be various ways of splitting an object according to a given rule and, moreover, some of them may yield the same result. We denote the collection of all possibilities by $\langle\varGamma\rangle=\left\{(\varGamma'',\varGamma')\right\}$ and for brevity write
\begin{eqnarray}
\varGamma\leadsto (\varGamma'',\varGamma') \in \langle\varGamma\rangle\ .
\end{eqnarray}
Note that strictly $\langle\varGamma\rangle$ is a multiset, \textit{i.e.} it is like a set but with arbitrary repetitions of elements allowed. Hence, in order not to overlook any of the decompositions, some of which may be the same, we should use a more appropriate notation employing the notion of a disjoint union, denoted by $\biguplus$, and write
\begin{eqnarray}
\langle\varGamma\rangle=\biguplus_{\substack{\text{decompositions}\\\varGamma\leadsto(\varGamma'',\varGamma')}}\left\{(\varGamma'',\varGamma')\right\}\ .
\end{eqnarray}
The concept of decomposition is quite general at this point and its further development obviously depends on the choice of the rule. One usually supplements this construction with additional constraints. Below we discuss some natural conditions one might expect from a decomposition rule. \vspace{0.2cm}
\begin{enumerate}
\item[\texttt{(0)}] {\textbf{\emph{Finiteness.}} It is reasonable to assume that an object decomposes in a finite number of ways, \textit{i.e.} for each $\varGamma$ the multiset $\langle\varGamma\rangle$ is finite.\vspace{0.2cm}}

\item[\texttt{(1)}] {\textbf{\emph{Triple decomposition.}} Decomposition into pairs naturally extends to splitting  an object into three pieces $\varGamma\leadsto (\varGamma_3,\varGamma_2,\varGamma_1)$. An obvious way to carry out the multiple splitting is by applying the same procedure repeatedly, \textit{i.e.} decomposing one of the components obtained in the preceding step. However, following this prescription one usually expects that the result does not depend on the choice of the component it is applied to. In other words, we require that we  end up with the same collection of triple decompositions when splitting $\varGamma\leadsto (\varGamma'',\varGamma_1)$ and then splitting the left component $\varGamma''\leadsto (\varGamma_3,\varGamma_2)$, \textit{i.e.}
\begin{eqnarray}\label{TripleDecomp1}
\varGamma\leadsto(\varGamma'',\varGamma_1)\leadsto(\varGamma_3,\varGamma_2,\varGamma_1)\ ,
\end{eqnarray}
as in the case when starting with $\varGamma\leadsto (\varGamma_3,\varGamma')$ and then splitting the right component $\varGamma'\leadsto (\varGamma_2,\varGamma_1)$, \textit{i.e.}
\begin{eqnarray}\label{TripleDecomp2}
\varGamma\leadsto(\varGamma_3,\varGamma')\leadsto(\varGamma_3,\varGamma_2,\varGamma_1)\ .
\end{eqnarray}
This condition can be seen as the co-associativity property for decomposition, and in explicit form boils down to the following equality:
\begin{eqnarray}\label{TripleDecomp}
\!\!\!\!\!\!\biguplus_{\substack{(\varGamma'',\varGamma_1)\in\langle\varGamma\rangle\\(\varGamma_3,\varGamma_2)\in\langle\varGamma''\rangle}}\!\!\!\left\{(\varGamma_3,\varGamma_2,\varGamma_1)\right\}\ =\!\!\!\!\biguplus_{\substack{(\varGamma_3,\varGamma')\in\langle\varGamma\rangle\\(\varGamma_2,\varGamma_1)\in\langle\varGamma'\rangle}}\!\!\!\left\{(\varGamma_3,\varGamma_2,\varGamma_1)\right\}.
\end{eqnarray}
The above procedure straightforwardly extends to splitting into multiple pieces $\varGamma\leadsto(\varGamma_n,...\,,\varGamma_1)$. Clearly, the condition of Eq.~(\ref{TripleDecomp}) entails the analogous property for multiple decompositions.
\vspace{0.2cm}}

\item[\texttt{(2)}] {\textbf{\emph{Void object.}}
Often, in a class there exists a sort of a void (or empty - we use both terms synonymously) element \O, such that objects decompose in a trivial way. It should have the the property that any object $\varGamma\neq\text{\O}$ splits into a pair containing either \O\ or $\varGamma$ in two ways only: \begin{eqnarray}
\varGamma\leadsto (\text{\O},\varGamma)\ \ \ \ \text{and}\ \ \ \  \varGamma\leadsto (\varGamma,\text{\O})\ ,
\end{eqnarray}
and $\text{\O}\leadsto (\text{\O},\text{\O})$. Clearly, if \O\ exists, it is unique.\vspace{0.2cm}}

\item[\texttt{(3)}] {\textbf{\emph{Symmetry.}} For some rules the order between components in decompositions  is immaterial, \textit{i.e.} the rule allows for an exchange $(\varGamma',\varGamma'')\longleftrightarrow (\varGamma'',\varGamma')$. In this case the following symmetry condition holds
\begin{eqnarray}\label{symmetry}
(\varGamma',\varGamma'')\in\langle\varGamma\rangle\ \Longleftrightarrow\ (\varGamma'',\varGamma')\in\langle\varGamma\rangle\ ,
\end{eqnarray}
and the multiplicities of $(\varGamma',\varGamma'')$ and $(\varGamma'',\varGamma')$ in $\langle\varGamma\rangle$ are the same.
\vspace{0.2cm}}

\item[\texttt{(4)}] {\textbf{\emph{Composition--decomposition compatibility.}}
Suppose that in addition to decomposition we also have a well-defined notion of composition of objects in the class. We denote the multiset comprising all possible compositions of $\varGamma_2$ with $\varGamma_1$  by $\varGamma_2\plug{\ }\varGamma_1$, \textit{e.g.} for the Heisenberg-Weyl diagrams we have
\begin{eqnarray}\label{plug}
\varGamma_2\plug{\ }\varGamma_1=\biguplus_{m\in\varGamma^{\stackrel{-}{}}_2\matchsubs{\ }\varGamma^{\stackrel{+}{}}_1} \varGamma_2\plug{m}\varGamma_1\ .
\end{eqnarray}
Now, given a pair of objects $\varGamma_2$ and $\varGamma_1$, we may think of two consistent decomposition schemes which involve composition. We can either start by composing them together $\varGamma_2\plug{}\varGamma_1$ and then splitting all resulting objects into pieces, or first  decompose each of them separately into $\langle\varGamma_2\rangle$ and $\langle\varGamma_1\rangle$ and then compose elements of both sets in a component-wise manner. One may require that the outcomes are the same no matter which way the procedure goes. Hence, a formal description of compatibility comes down to the equality:
\begin{eqnarray}\label{compatibility}
\biguplus_{\varGamma\in\varGamma_2\plug{}\varGamma_1}\langle\varGamma\rangle\ =\!\!\!\biguplus_{\substack{(\varGamma_2'',\varGamma_2')\in\langle\varGamma_2\rangle\\(\varGamma_1'',\varGamma_1')\in\langle\varGamma_1\rangle}}\!\!(\varGamma_2''\plug{}\varGamma_1'')\times(\varGamma_2'\plug{}\varGamma_1')\ .
\end{eqnarray}
We remark that this property indicates that the void object \O\ of condition \texttt{(2)} is the same as the neutral element for composition.
\vspace{0.2cm}}

\item[\texttt{(5)}] {\textbf{\emph{Finiteness of multiple decompositions.}}
Recall the process of multiple decompositions $\varGamma\leadsto(\varGamma_n,...\varGamma_1)$ constructed in condition \texttt{(1)} and observe that one may extend the number of components to any $n\in\mathbb{N}$.  However, if one considers only nontrivial decompositions which do not contain void components \O\ it is often the case that the process terminates after a finite number of steps.
In other words, for each $\varGamma$ there exists $N\in\mathbb{N}$ such that
\begin{eqnarray}\label{FinMultDecomp}
\left\{\varGamma\leadsto(\varGamma_n,...\varGamma_1): \varGamma_n,...,\varGamma_1\neq\text{\O}\right\}=\emptyset
\end{eqnarray}
for $n>N$.
In practice, objects usually carry various characteristics counted by natural numbers, \textit{e.g.} the number of elements they are built from. Then, if the decomposition rule decreases such a characteristic in each of the components in a nontrivial splitting, it inevitably exhausts and then the condition of Eq.~(\ref{FinMultDecomp}) is automatically fulfilled.}
\end{enumerate}

Having discussed the above quite general conditions expected from a reasonable decomposition rule we are now in a position to return to the realm of algebra.
We have already seen in Section~\ref{AlgebraHWDiag} how the notion of composition induces a multiplication which endows the class of Heisenberg-Weyl diagrams with the structure of an algebra, see Theorem~\ref{Diag-AAU}. Following this route we now employ the concept of decomposition to introduce the structure of a Hopf algebra in $\mathcal{G}$. A central role in the construction will be played by the  three mappings given below.

Let us consider a linear mapping $\Delta:\mathcal{G}\longrightarrow\mathcal{G}\otimes\mathcal{G}$ defined on the basis elements as a sum of possible splittings, \textit{i.e.}
\begin{eqnarray}\label{delta}
\Delta(\varGamma)=\sum_{(\varGamma',\varGamma'')\in\langle\varGamma\rangle}\varGamma'\otimes\varGamma''\ .
\end{eqnarray}
Note, that although all coefficients in Eq.~(\ref{delta}) are equal to one, some terms in the sum may appear several times. This is because elements in the multiset $\langle\varGamma\rangle$ may repeat and the numbers counting their multiplicities are sometimes called section coefficients \cite{JoniRota}. Observe that the sum is well defined as long the number of decompositions is finite, \textit{i.e.} condition \texttt{(0)} is satisfied.

We also make use of a linear mapping $\varepsilon:\mathcal{G}\longrightarrow\mathbb{K}$ which extracts the coefficient of the  void element  \O. It is defined on the basis elements by:
\begin{eqnarray}\label{epsilon}
\varepsilon(\varGamma)=\left\{
\begin{array}{l}
1 \text{\ \ \ \ if\ \ \ }\varGamma=\text{\O}\ ,\\
0  \text{\ \ \ \ otherwise\ . }
\end{array}\right.
\end{eqnarray}

Finally, we need a linear mapping $S:\mathcal{G}\longrightarrow\mathcal{G}$ defined by the formula
\begin{eqnarray}\label{antipode}
S(\varGamma)=\sum_{\substack{\varGamma\leadsto(\varGamma_n,...,\varGamma_1) \\ \varGamma_n,...,\varGamma_1\neq\emptyset}}(-1)^n\ \varGamma_n*...*\varGamma_1\ ,
\end{eqnarray}
for $\varGamma\neq\text{\O}$ and $S(\text{\O})=\text{\O}$. Note that it is an alternating sum over products of nontrivial multiple decompositions of an object. Clearly, if the condition \texttt{(5)} holds the sum is finite and $S$ is well defined.

The mappings $\Delta$, $\varepsilon$ and $S$, built upon a reasonable decomposition procedure, provide $\mathcal{G}$ with a rich algebraic structure as summarized in the following lemma (for the proofs see Appendix~\ref{Appendix-Hopf}):

\begin{lemma}[Decomposition and Hopf algebra] \label{bi-algebra}\ \vspace{-0.4cm}\\
\begin{enumerate}
\item[\textbf{(i)}]{If the conditions \emph{\texttt{(0)}}, \emph{\texttt{(1)}} and \emph{\texttt{(2)}} are satisfied, the mappings $\Delta$ and $\varepsilon$ defined in Eqs.~(\ref{delta}) and (\ref{epsilon}) are the co-product and co-unit in the algebra $\mathcal{G}$.  The co-algebra $(\mathcal{G},\Delta,\varepsilon)$ thus defined is co-commutative, provided condition \emph{\texttt{(3)}} is fulfilled.\vspace{0.1cm}}
\item[\textbf{(ii)}]{In addition, if  condition \emph{\texttt{(4)}} holds we have a genuine bi-algebra structure $(\mathcal{G},+,*,\text{\emph{\O}},\Delta,\varepsilon)$.\vspace{0.1cm}}
\item[\textbf{(iii)}]{Finally, under condition \emph{\texttt{(5)}} we establish a Hopf algebra structure $(\mathcal{G},+,*,\text{\emph{\O}},\Delta,\varepsilon,S)$ with the antipode $S$ defined in Eq.~(\ref{antipode}).}
\end{enumerate}
\end{lemma}

We remark that the above discussion is applicable to a wide range of combinatorial classes and decomposition rules which we have thus far  left unspecified. Below, we apply these concepts to the class of Heisenberg-Weyl diagrams.



\subsection{Hopf algebra of Heisenberg-Weyl diagrams}

In this Section, we  provide an explicit decomposition rule for the Heisenberg-Weyl diagrams satisfying all the conditions discussed in Section~\ref{BasicNotions}. In this way we  complete the whole picture by introducing a Hopf algebra structure on $\mathcal{G}$.

\begin{figure*}[th]
\begin{center}
\resizebox{\textwidth}{!}{\includegraphics{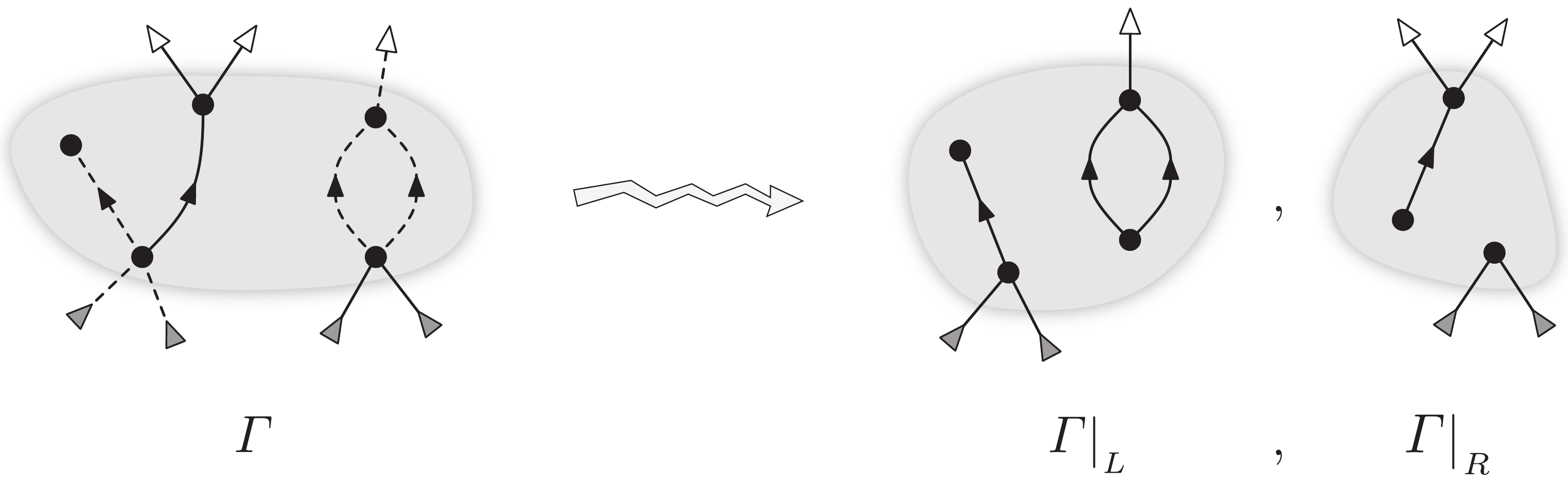}}
\caption{\label{Diagram-decomp} An example of diagram decomposition $\varGamma\leadsto (\left.\varGamma\right|_L,\left.\varGamma\right|_R)$. The choice of edges $L\subset E_\varGamma$ inducing the diagram $\left.\varGamma\right|_L$ is depicted on the left diagram as dashed lines.}
\end{center}
\end{figure*}

We start by observing that for a given Heisenberg-Weyl graph $\varGamma$, each subset of its edges $L\subset E_\varGamma$ induces a subgraph $\left.\varGamma\right|_L$ which is defined by restriction of the head and tail functions to the subset $L$.  Likewise, the remaining part of the edges $R=E_\varGamma-L$ gives rise to a subgraph $\left.\varGamma\right|_R$. Clearly, the results are again Heisenberg-Weyl graphs. Thus, by considering ordered partitions of the set of edges into two subsets $L+R=E_\varGamma$, \textit{i.e.} $L\cup R=E_\varGamma$ and $L\cap R=\emptyset$, we end up with pairs of disjoint graphs $(\left.\varGamma\right|_L,\left.\varGamma\right|_R)$. This suggests the following definition:

\pagebreak

\begin{definition}[Diagram Decomposition]\label{HWDiag-decomp} \ \vspace{0.1cm}\\
A decomposition of a Heisenberg-Weyl diagram $\varGamma$ is any splitting $(\varGamma_L,\varGamma_R)$ induced by an ordered partition of its lines $L+R=E_\varGamma$. Hence, the multiset $\langle\varGamma\rangle$ comprising all possible decompositions can be indexed by the set of ordered double partitions $\{(L,R):\ L+R=E_\varGamma\}$, and we have
\begin{eqnarray}
\langle\varGamma\rangle=\biguplus_{L+R=E_\varGamma}\left\{(\left.\varGamma\right|_L,\left.\varGamma\right|_R)\right\}\ .
\end{eqnarray}
\end{definition}
The graphical picture is clear: the decomposition of a diagram $\varGamma\leadsto (\left.\varGamma\right|_L,\left.\varGamma\right|_R)$ is defined by the choice of lines $L\subset E_\varGamma$, which taken out make up the first component of the pair whilst the remainder induced by $R=E_\varGamma-L$ constitutes the second one. (See the illustration in Fig.~\ref{Diagram-decomp}.)

We observe that the enumeration of all decompositions of a diagram $\varGamma$ is straightforward since the multiset $\langle\varGamma\rangle$ can be indexed by subsets of $E_\varGamma$. Because $|E_\varGamma|=|\varGamma|$, explicit counting gives
$|\langle\varGamma\rangle|=\sum_i \binom{|\varGamma|}{i}=2^{|\varGamma|}$. This simple observation can be generalized to calculate the number of decompositions $(\left.\varGamma\right|_L,\left.\varGamma\right|_R)\in\langle\varGamma\rangle$ in which the first component has $i$ outgoing, $j$ incoming and $k$ inner lines, \textit{i.e.} $|\left.\varGamma\right|_L^{\stackrel{+}{}}|=i, |\left.\varGamma\right|_L^{\stackrel{-}{}}|=j,|\left.\varGamma\right|_L^{\stackrel{0}{}}|=k$.  Accordingly, the enumeration reduces to the choice of $i$, $j$ and $k$ lines out of the sets $\varGamma^{\stackrel{+}{}}$, $\varGamma^{\stackrel{-}{}}$ and $\varGamma^{\stackrel{0}{}}$ respectively, which gives
\begin{eqnarray}\nonumber
\left|\left\{\left(\left.\varGamma\right|_L,\left.\varGamma\right|_R\right)\in\langle\varGamma\rangle: \substack{|\left.\varGamma\right|_L^{\stackrel{+}{}}|=i\\|\left.\varGamma\right|_L^{\stackrel{-}{}}|=j\\|\left.\varGamma\right|_L^{\stackrel{0}{}}|=k}\right\}\right|=\binom{|\varGamma^{\stackrel{+}{}}|}{i}\binom{|\varGamma^{\stackrel{-}{}}|}{j}\binom{|\varGamma^{\stackrel{0}{}}|}{k}.\!\!\!\!\!\!\!\\\label{decomp-count}
\end{eqnarray}
Of course, the second component $\left.\varGamma\right|_R$ is always determined by the first one $\left.\varGamma\right|_L$ and hence the number of its outgoing, incoming and inner lines is given by
\begin{eqnarray}\nonumber
|\left.\varGamma\right|_R^{\stackrel{+}{}}|&=&|\varGamma^{\stackrel{+}{}}|-i\ ,\\\label{decomp-ijk}
|\left.\varGamma\right|_R^{\stackrel{-}{}}|&=&|\varGamma^{\stackrel{-}{}}|-j\ ,\\\nonumber
|\left.\varGamma\right|_R^{\stackrel{0}{}}|&=&|\varGamma^{\stackrel{0}{}}\,|-k\ .
\end{eqnarray}

Having explicitly defined the notion of diagram decomposition, one may check that it satisfies conditions \texttt{(1)} - \texttt{(5)} of Section~\ref{BasicNotions}; for the proofs see Appendix~\ref{PropDiagDecomp}.
In this context Eq.~(\ref{delta}) defining the co-product in the algebra $\mathcal{G}$ takes the form
\begin{eqnarray}\label{delta-diag}
\Delta(\varGamma)=\sum_{L+R=E_\varGamma}\left.\varGamma\right|_L\otimes\left.\varGamma\right|_R\ ,
\end{eqnarray}
and the antipode of Eq.~(\ref{antipode}) may be rewritten  as
\begin{eqnarray}\label{antipode-HWDiag}
S(\varGamma)=\sum_{\substack{A_n+...+A_1=E_\varGamma \\ A_n,...,A_1\neq\emptyset}}(-1)^n\ \left.\varGamma\right|_{A_n}*...*\left.\varGamma\right|_{A_1}\ .
\end{eqnarray}
for $\varGamma\neq\text{\O}$ and $S(\text{\O})=\text{\O}$.
Therefore, referring to Lemma~\ref{bi-algebra}, we supplement Theorem~\ref{Diag-AAU} by the following result:
\begin{theorem}[Hopf algebra of Diagrams] \label{HopfAlgDiag} \ \vspace{0.1cm}\\
The algebra of Heisenberg-Weyl diagrams $\mathcal{G}$ has a Hopf algebra structure $(\mathcal{G},+,*,\text{\emph{\O}},\Delta,\varepsilon,S)$ with (co-commutative) co-product, co-unit and antipode as defined in Eqs.~(\ref{delta-diag}), (\ref{epsilon}) and (\ref{antipode-HWDiag}) respectively.
\end{theorem}

The algebra of Heisenberg-Weyl diagrams $\mathcal{G}$ was shown to be directly related to the algebra $\mathcal{U}(\mathcal{L}_{\mathcal{H}})$ through the forgetful mapping $\varphi$ which preserves algebraic operations as explained in Theorem~\ref{Forgetful-mapping}. Here, however, in the context of Theorem~\ref{HopfAlgDiag} the algebra $\mathcal{G}$ is additionally equipped with a co-product, co-unit and antipode. Since $\mathcal{U}(\mathcal{L}_{\mathcal{H}})$ is also a Hopf algebra, it is natural to ask whether this extra structure is preserved by the morphism  $\varphi$ of Eq.~(\ref{phi}). It turns out that indeed it is also preserved, and one can augment Theorem~\ref{Forgetful-mapping} in the following way (for the proof see Appendix~\ref{Appendix-Morphism}):

\begin{theorem}[Hopf algebra morphism $\varphi$] \label{Hopf-alg-morph} \ \vspace{0.1cm}\\
The forgetful mapping $\varphi:\mathcal{G}\longrightarrow\mathcal{U}(\mathcal{L}_{\mathcal{H}})$ defined in Eq.~(\ref{phi}) is a Hopf algebra morphism.
\end{theorem}

In this way, we have extended the results of Section~\ref{HWDiag} to encompass the Hopf algebra structure of the enveloping algebra $\mathcal{U}(\mathcal{L}_\mathcal{H})$.
This completes the picture of the algebra of Heisenberg-Weyl diagrams $\mathcal{G}$ as a combinatorial model which captures all the relevant properties of the algebras $\mathcal{H}$ and $\mathcal{U}(\mathcal{L}_\mathcal{H})$.

\section{Conclusions}

The development of concrete models in physics often provides a means of  understanding abstract algebraic constructs in a more natural way. This  appears to  be particularly valuable in the realm of Quantum Theory, where the abstract formalism is far from intuitive. In this respect, the  combinatorial perspective seems to provide a promising approach, and as such has become a blueprint for much contemporary research.
For example, recent work in perturbative Quantum Field Theory (pQFT) has shown the value of analyzing the algebraic structure of a diagrammatic approach, in the case of pQFT, that of the Feynman diagrams \cite{KreimerBook}.  The present work differs from that discussing  pQFT in several respects. Standard non-relativistic second-quantized quantum theory, in which context the present study is firmly based, does not suffer from the singularities which plague pQFT. As a consequence, well-understood  procedures will, at least in principle, suffice to analyze models based on non-relativistic quantum theory. Nevertheless, the value both of a diagrammatic approach -- even in the non-relativistic case -- as well as an analysis of the underlying algebraic structure -- can only lead to a deeper understanding of the theory. In this note we described perhaps the most basic structure of quantum theory, that involving a single mode second-quantized theory\footnote{One does not expect that the extension to several {\em commuting} modes would introduce additional complication.}. In spite of this simple model, the underlying algebraic structure proves to be surprisingly rich\footnote{Of course, this is not identical to the Connes-Kreimer algebra arising in pQFT.}.

The standard commutation relation between a single  creation and annihilation operator of second-quantized quantum mechanics,  $a\,a^\dag-a^\dag a=I$, generates in a natural way  the Heisenberg-Weyl associative algebra $\mathcal{H}$, as well as  the  Heisenberg Lie algebra $\mathcal{L}_\mathcal{H}$ and its enveloping algebra $\mathcal{U}(\mathcal{L}_{\mathcal{H}})$.  We discussed these algebras, showing, {\it inter alia}, that  $\mathcal{U}(\mathcal{L}_{\mathcal{H}})$  can be endowed with a Hopf algebra structure, unlike $\mathcal{H}$. However, the main content of the current  work was the introduction of a combinatorial algebra $\mathcal{G}$ of graphs, arising from a diagrammatic representation of the creation-annihilation operator system. This algebra was shown to carry a natural Hopf structure. Further, it was proved that both $\mathcal{H}$ and $\mathcal{U}(\mathcal{L}_{\mathcal{H}})$ were homomorphic images of $\mathcal{G}$, in the latter case a true Hopf algebra homomorphism.

Apart from giving a concrete and visual representation of the $a$ and $a^\dag$ actions, the algebra $\mathcal{G}$ remarkably exhibits a finer structure than either  of the algebras $\mathcal{H}$ or  $\mathcal{U}(\mathcal{L}_{\mathcal{H}})$.   This ``fine graining'' of the effective actions of the creation-annihilation operators implies a richer structure for these actions, possibly leading to a deeper insight into this basic quantum mechanical system. Moreover, we should point out that the  diagrammatic model of the Heisenberg-Weyl algebra presented here is particularly suited to the methods of modern combinatorial analysis \cite{FlajoletBook,BergeronBook,AignerBook}; we intend to develop this aspect  in a forthcoming publication.

\section*{Acknowledgments}

We wish to thank Philippe Flajolet for important discussions on the subject.
Most of this research was carried out in the Mathematisches Forschungsinstitut Oberwolfach (Germany) and the Laboratoire d'Informatique de l'Universit\'e Paris-Nord in Villetaneuse (France) whose warm hospitality is greatly appreciated.
The authors acknowledge support from the Polish Ministry of Science and Higher Education grant no. N202 061434 and the Agence Nationale de la Recherche under programme no. ANR-08-BLAN-0243-2.

\newpage

\appendix

\section*{Appendixes}

\section{Associativity of multiplication in $\mathcal{G}$}\label{Appendix-Associativity}

We prove associativity of the multiplication defined in Eq.~(\ref{multiplication-def}). From bilinearity, we  need only check it for the basis elements, \textit{i.e.}
\begin{eqnarray}
\varGamma_3*(\varGamma_2*\varGamma_1)=(\varGamma_3*\varGamma_2)*\varGamma_1\ .
\end{eqnarray}
Written explicitly, the left- and right-hand sides of this equation take the form
\begin{eqnarray}\label{A1}
\varGamma_3*(\varGamma_2*\varGamma_1)=\sum_{m'}\sum_{m_{21}}\ \varGamma_3\plug{m'}(\varGamma_2\plug{m_{21}}\varGamma_1)
\end{eqnarray}
where $m'\in\varGamma^{\stackrel{-}{}}_3\match{\ }(\varGamma_2\plug{m_{21}}\varGamma_1)^{\stackrel{+}{}}$ and $m_{21}\in\varGamma^{\stackrel{-}{}}_2\match{\ }\varGamma^{\stackrel{+}{}}_1$, whilst
\begin{eqnarray}\label{A2}
(\varGamma_3*\varGamma_2)*\varGamma_1=\sum_{m_{32}}\sum_{m''}\ (\varGamma_3\plug{m_{32}}\varGamma_2)\plug{m''}\varGamma_1
\end{eqnarray}
where $m_{32}\in\varGamma^{\stackrel{-}{}}_3\match{\ }\varGamma^{\stackrel{+}{}}_2$ and $m''\in(\varGamma_3\plug{m_{32}}\varGamma_2)^{\stackrel{-}{}}\match{\ }\varGamma^{\stackrel{+}{}}_1$.

\begin{figure}[h]
\begin{center}
\resizebox{0.55\textwidth}{!}{\includegraphics{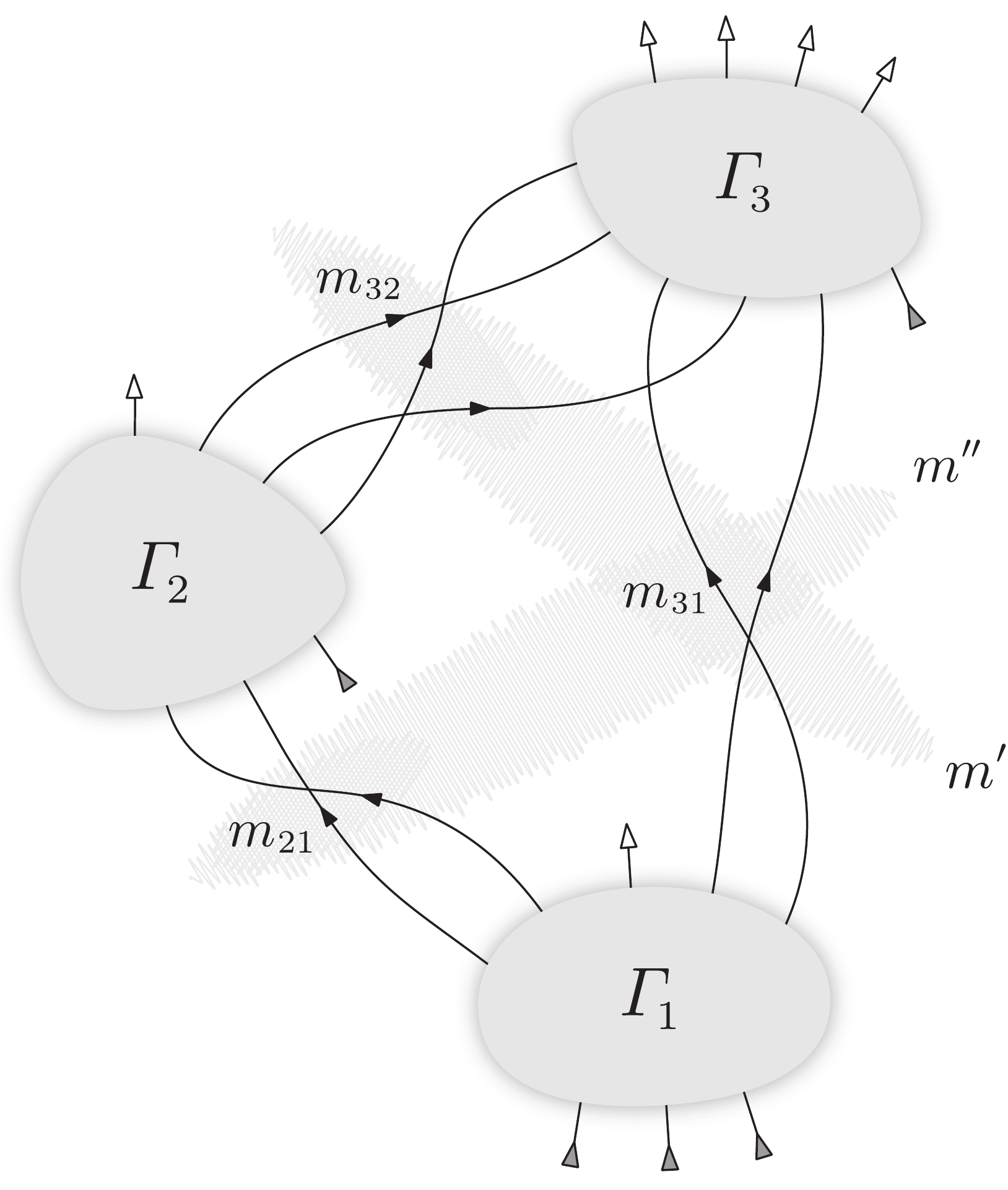}}
\caption{\label{Diagram-associativity-proof} Fine graining of the matchings $m'\in\varGamma^{\stackrel{-}{}}_3\match{\ }(\varGamma_2\plug{m_{21}}\varGamma_1)^{\stackrel{+}{}}$ and $m''\in(\varGamma_3\plug{m_{32}}\varGamma_2)^{\stackrel{-}{}}\match{\ }\varGamma^{\stackrel{+}{}}_1$ used in the proof of associativity of multiplication.}
\end{center}
\end{figure}

Consider the double sums in the above equations, indexed by $(m',m_{21})$ and $(m_{32},m'')$ respectively, and observe that there exists a one-to-one correspondence between their elements. We construct it by a fine graining of the matchings, see Fig.~\ref{Diagram-associativity-proof}, and define the following two mappings. The first one is
\begin{eqnarray}
(m',m_{21})\ \longrightarrow\ (m_{32},m'')\ ,
\end{eqnarray}
where
$m_{32}=m'\cap(\varGamma_3^{\stackrel{-}{}}\times\varGamma_2^{\stackrel{+}{}})$ and
$m''=m_{21}\cup(m'\cap(\varGamma_3^{\stackrel{-}{}}\times\varGamma_1^{\stackrel{+}{}}))$, and similarly the second one
\begin{eqnarray}
(m_{32},m'')\ \longrightarrow\ (m',m_{21})\ ,
\end{eqnarray}
with $m'=m_{32}\cup(m''\cap(\varGamma_3^{\stackrel{-}{}}\times\varGamma_1^{\stackrel{+}{}}))$ and
$m_{21}=m''\cap(\varGamma_2^{\stackrel{-}{}}\times\varGamma_1^{\stackrel{+}{}})$. Clearly, the mappings are inverses of each other, which ensures a one-to-one correspondence between elements of the double sums in Eqs.~(\ref{A1}) and (\ref{A2}). Moreover, the summands that are mapped  onto each other are equal, \textit{i.e.} the corresponding diagrams $\varGamma_3\plug{m'}(\varGamma_2\plug{m_{21}}\varGamma_1)$ and $(\varGamma_3\plug{m_{32}}\varGamma_2)\plug{m''}\varGamma_1$ are exactly the same. This completes the proof by showing equality of the right-hand sides of Eqs.~(\ref{A1}) and (\ref{A2}).

\pagebreak

\section{Forgetful morphism $\varphi$}\label{Appendix-Morphism}

In Theorems~\ref{Forgetful-mapping} and \ref{Hopf-alg-morph} we stated that the linear mapping $\varphi:\mathcal{G}\longrightarrow\mathcal{U}(\mathcal{L}_{\mathcal{H}})$ defined in Eq.~(\ref{phi}) was a Hopf algebra morphism. We now prove this statement.

We start by showing that $\varphi$ preserves multiplication in $\mathcal{G}$. From linearity it is enough to check for the basis elements that $\varphi(\varGamma_2*\varGamma_1)=\varphi(\varGamma_2)\,\varphi(\varGamma_1)$, which is verified in the following sequence of equalities:
\begin{eqnarray}
\varphi(\varGamma_2*\varGamma_1)&\stackrel{(\ref{multiplication-def})}{=}&\sum_{m\in\varGamma^{\stackrel{-}{}}_2\matchsubs{\  }\varGamma^{\stackrel{+}{}}_1}\varphi(\varGamma_2\plug{m}\varGamma_1)
=\sum_i\sum_{m\in\varGamma_2\matchsubs{i}\varGamma_1}\varphi(\varGamma_2\plug{m}\varGamma_1)\label{B1}\\\nonumber
&\stackrel{(\ref{phi-of-match})}{=}&\sum_i\sum_{m\in\varGamma^{\stackrel{-}{}}_2\matchsubs{i }\varGamma^{\stackrel{+}{}}_1}(a^\dag)^{\,|\varGamma_2^{\stackrel{+}{}}\!|+|\varGamma_1^{\stackrel{+}{}}\!|-i}\ a^{\,|\varGamma_2^{\stackrel{-}{}}\!|+|\varGamma_1^{\stackrel{-}{}}\!|-i}\ e^{\,|\varGamma_2^{\stackrel{0}{}}|+|\varGamma_1^{\stackrel{0}{}}|+i}\\
&=&\sum_i\ (a^\dag)^{\,|\varGamma_2^{\stackrel{+}{}}\!|+|\varGamma_1^{\stackrel{+}{}}\!|-i}\ a^{\,|\varGamma_2^{\stackrel{-}{}}\!|+|\varGamma_1^{\stackrel{-}{}}\!|-i}\ e^{\,|\varGamma_2^{\stackrel{0}{}}|+|\varGamma_1^{\stackrel{0}{}}|+i}\sum_{m\in\varGamma^{\stackrel{-}{}}_2\matchsubs{i}\varGamma^{\stackrel{+}{}}_1}1\label{B3}\\\nonumber
&\stackrel{(\ref{num-diag-match})}{=}&\sum_i\binom{|\varGamma_2^{\stackrel{-}{}}\!|}{i}\binom{|\varGamma_1^{\stackrel{+}{}}\!|}{i}\ i!\ (a^\dag)^{\,|\varGamma_2^{\stackrel{+}{}}\!|+|\varGamma_1^{\stackrel{+}{}}\!|-i}\ a^{\,|\varGamma_2^{\stackrel{-}{}}\!|+|\varGamma_1^{\stackrel{-}{}}\!|-i}\ e^{\,|\varGamma_2^{\stackrel{0}{}}|+|\varGamma_1^{\stackrel{0}{}}|+i}\\\nonumber
&\stackrel{(\ref{StrConst})}{=}&\left((a^\dag)^{\,|\varGamma_2^{\stackrel{+}{}}\!|}\ a^{\,|\varGamma_2^{\stackrel{-}{}}\!|}\ e^{\,|\varGamma_2^{\stackrel{0}{}}|}\right)\left((a^\dag)^{\,|\varGamma_1^{\stackrel{+}{}}\!|}\ a^{\,|\varGamma_1^{\stackrel{-}{}}\!|}\ e^{\,|\varGamma_1^{\stackrel{0}{}}|}\right)=\varphi(\varGamma_2)\,\varphi(\varGamma_1)\ .
\end{eqnarray}
In the above derivation the main trick in Eq.~(\ref{B1}) consists of splitting the set of diagram matchings into disjoint subsets according to the number of connected lines, \textit{i.e.} $\varGamma^{\stackrel{-}{}}_2\match{\ }\varGamma^{\stackrel{+}{}}_1=\bigcup_i\varGamma^{\stackrel{-}{}}_2\match{i}\varGamma^{\stackrel{+}{}}_1$. Then, observing  that the summands in Eq.~(\ref{B3}) do not depend on $m\in\varGamma^{\stackrel{-}{}}_2\match{i}\varGamma^{\stackrel{+}{}}_1$, we may execute explicitly one of the sums counting elements in $\varGamma^{\stackrel{-}{}}_2\match{i}\varGamma^{\stackrel{+}{}}_1$ with the help of Eq.~(\ref{num-diag-match}).

We also need to show that the co-product, co-unit and antipode are preserved by $\varphi$.
This means that when proceeding via the mapping $\varphi$ from $\mathcal{G}$ to $\mathcal{U}(\mathcal{L}_{\mathcal{H}})$ one can use the co-product, co-unit and antipode in either of the algebras and obtain the same result \textit{i.e.}
\begin{eqnarray}\label{phi-co-product}
(\varphi\otimes\varphi)\circ\Delta&=&\Delta\circ\varphi\ ,\\\label{phi-co-unit}
\varepsilon&=&\varepsilon\circ\varphi\ ,\\\label{phi-antipode}
\varphi\circ S&=&S\circ\varphi\ ,
\end{eqnarray}
where $\Delta$, $\varepsilon$ and $S$ on the left-hand sides act in $\mathcal{G}$ whilst on the right-hand sides in $\mathcal{U}(\mathcal{L}_{\mathcal{H}})$. The proof of Eq.~(\ref{phi-co-product}) rests upon the counting formula in Eq.~(\ref{decomp-count}) and the observation of Eq.~(\ref{decomp-ijk}), which justify the following equalities
\begin{eqnarray}\nonumber
(\varphi\otimes\varphi)\circ\Delta\,(\varGamma)&\!\!\!\!=&\!\!\!\sum_{L+R=E_\varGamma}\varphi\left(\left.\varGamma\right|_L\right)\otimes\varphi\left(\left.\varGamma\right|_R\right)
=\sum_{L\subset E_\varGamma}\varphi\left(\left.\varGamma\right|_L\right)\otimes\varphi\left(\left.\varGamma\right|_{E_\varGamma-L}\right)\\\nonumber
&\!\!\!\!\stackrel{(\ref{decomp-count}),(\ref{decomp-ijk})}{=}&\!\!\!\sum_{i,j,k}\binom{|\varGamma^{\stackrel{+}{}}|}{i}\binom{|\varGamma^{\stackrel{-}{}}|}{j}\binom{|\varGamma^{\stackrel{0}{}}|}{k}\ a^{\dag\,i}\ a^j\ e^k\otimes a^{\dag\,|\varGamma^{\stackrel{+}{}}|-i}\ a^{|\varGamma^{\stackrel{-}{}}|-j}\ e^{|\varGamma^{\stackrel{0}{}}|-k}\\\nonumber&\!\!\!\!
\stackrel{(\ref{Uco-product})}{=}&\!\!\!\Delta\circ\varphi\,(\varGamma)\ .
\end{eqnarray}
Eq.~(\ref{phi-co-unit}) is readily verified by comparing Eqs.~(\ref{epsilon}) and (\ref{Uco-unit}).
Eq.~(\ref{phi-antipode}) is similarly checked, as the structure of Eq.~(\ref{antipode}) faithfully transfers via morphism into the analogous general formula for the antipode in the graded Hopf algebras (see \cite{AbeBook,CartierHopfPrimer}), the latter of course reproducing Eq.~(\ref{Uantipode}) in the case of Lie algebras.

\section{From decomposition to Hopf algebra}\label{Appendix-Hopf}

In order to prove Lemma~\ref{bi-algebra} we should check in part \emph{(i)} co-associativity of the co-product $\Delta$ and properties of the co-unit $\varepsilon$, in part \emph{(ii)} show that the mappings $\Delta$ and $\varepsilon$ preserve multiplication in $\mathcal{G}$, and for part \emph{(iii)} verify the defining properties of the antipode $S$.

\subsection*{\emph{(i)} Co-algebra}
The co-product $\Delta:\mathcal{G}\longrightarrow\mathcal{G}\otimes\mathcal{G}$ is co-associative
if the following equality holds
\begin{eqnarray} \label{coassoc}
(\Delta\otimes Id)\circ\Delta=(Id\otimes\Delta)\circ\Delta\ .
\end{eqnarray}
Since $\Delta$ defined in Eq.~(\ref{delta}) is linear it is enough to check (\ref{coassoc}) for the basis elements $\varGamma$. Accordingly, the left-hand side takes the form
\begin{eqnarray}\label{associativity1}
(\Delta\otimes Id)\circ\Delta\, (\varGamma)=(Id\otimes\Delta)\sum_{(\varGamma_1,\varGamma'')\in\langle\varGamma\rangle}\varGamma_1\otimes\varGamma''=\sum_{\substack{(\varGamma_1,\varGamma'')\in\langle\varGamma\rangle\\(\varGamma_2,\varGamma_3)\in\langle\varGamma''\rangle}}\varGamma_1\otimes\varGamma_2\otimes\varGamma_3\ ,
\end{eqnarray}
whereas the right-hand side is
\begin{eqnarray}\label{associativity2}
(Id\otimes \Delta)\circ\Delta\, (\varGamma)=(\Delta\otimes Id)\sum_{(\varGamma',\varGamma_3)\in\langle\varGamma\rangle}\varGamma'\otimes\varGamma_3=\sum_{\substack{(\varGamma',\varGamma_3)\in\langle\varGamma\rangle\\(\varGamma_1,\varGamma_2)\in\langle\varGamma'\rangle}}\varGamma_1\otimes\varGamma_2\otimes\varGamma_3\ .
\end{eqnarray}
If  condition \texttt{(1)} of Section~\ref{BasicNotions} holds, the property Eq.~(\ref{TripleDecomp}) asserts equality of the right-hand sides of Eqs.~(\ref{associativity1}) and (\ref{associativity2}) and the co-product defined in Eq.~(\ref{delta}) is co-associative.

By definition, the co-unit $\varepsilon: \mathcal{G}\longrightarrow \mathbb{K}$  should satisfy the equalities
\begin{eqnarray}\label{co-unit-def}
(\varepsilon\otimes Id)\circ\Delta=Id=(Id\otimes\varepsilon)\circ\Delta \ ,
\end{eqnarray}
where the identification $\mathbb{K}\otimes\mathcal{G}=\mathcal{G}\otimes\mathbb{K}=\mathcal{G}$ is implied.
We check the first one for the basis elements $\varGamma$ by direct calculation:
\begin{eqnarray}\nonumber
(\varepsilon\otimes Id)\circ\Delta\,(\varGamma)&=&(\varepsilon\otimes Id)\sum_{(\varGamma_1,\varGamma_2)\in\langle\varGamma\rangle}\varGamma_1\otimes\varGamma_2\\\label{epssum}
&=&\sum_{(\varGamma_1,\varGamma_2)\in\langle\varGamma\rangle}\varepsilon(\varGamma_1)\otimes\varGamma_2\\\nonumber
&=&1\otimes\varGamma=\varGamma=Id\,(\varGamma)\ .
\end{eqnarray}
Note that we have applied  condition \texttt{(2)} of Section~\ref{BasicNotions} by taking all terms in the sum Eq.~(\ref{epssum}) equal to zero except the unique decomposition $(\text{\O},\varGamma)$ picked up by $\varepsilon$ as defined in Eq.~(\ref{epsilon}). The identification $1\otimes\varGamma=\varGamma$ completes the proof of the first equality in Eq.~(\ref{co-unit-def}); verification of the second one is analogous.

Co-commutativity of the co-product $\Delta$ under the condition \texttt{(3)} is straightforward since from Eq.~(\ref{symmetry}) we have
\begin{eqnarray}\nonumber
\Delta(\varGamma)=\sum_{(\varGamma',\varGamma'')\in\langle\varGamma\rangle}\varGamma'\otimes\varGamma''=\sum_{(\varGamma',\varGamma'')\in\langle\varGamma\rangle}\varGamma''\otimes\varGamma'\ .
\end{eqnarray}

\subsection*{\emph{(ii)} Bi-algebra}
The structure of a bi-algebra results whenever the co-product $\Delta:\mathcal{G}\otimes\mathcal{G}\longrightarrow\mathcal{G}$ and co-unit $\varepsilon:\mathcal{G}\longrightarrow\mathbb{K}$ of the co-algebra are compatible with multiplication in $\mathcal{G}$. Thus, we need to verify for basis elements $\varGamma_1$ and $\varGamma_2$ that
\begin{eqnarray}\label{delta-morph}
\Delta\,(\varGamma_2*\varGamma_1)&=&\Delta\,(\varGamma_2)*\Delta\,(\varGamma_1)\ ,
\end{eqnarray}
with component-wise multiplication in the tensor product $\mathcal{G}\otimes\mathcal{G}$ on the right-hand-side, and
\begin{eqnarray}\label{epsilon-morph}
\varepsilon\,(\varGamma_2*\varGamma_1)&=&\varepsilon\,(\varGamma_2)\ \varepsilon\,(\varGamma_1)\ ,
\end{eqnarray}
with terms on the right-hand-side multiplied in $\mathbb{K}$.

We check Eq.~(\ref{delta-morph}) directly by expanding both  sides using the definitions of Eqs.~(\ref{multiplication-def}), (\ref{plug}) and (\ref{delta}). Accordingly, the left-hand-side takes the form
\begin{eqnarray}\label{DM1}
\Delta\,(\varGamma_2*\varGamma_1)=\sum_{\varGamma\in\varGamma_2\plug{}\varGamma_1}\Delta\,(\varGamma)
=\sum_{\varGamma\in\varGamma_2\plug{}\varGamma_1}\ \ \sum_{(\varGamma'',\varGamma')\in\langle\varGamma\rangle}\varGamma''\otimes\varGamma'\ ,
\end{eqnarray}
while the right-hand side is
\begin{eqnarray}\nonumber
\Delta\,(\varGamma_2)*\Delta\,(\varGamma_1)&=&\sum_{\substack{(\varGamma_2'',\varGamma_2')\in\langle\varGamma_2\rangle\\(\varGamma_1'',\varGamma_1')\in\langle\varGamma_1\rangle}}\underbrace{(\varGamma_2''\otimes\varGamma_2')*(\varGamma_1''\otimes\varGamma_1')}_{(\varGamma_2''*\varGamma_1'')\otimes(\varGamma_2'*\varGamma_1')}\\
&=&\sum_{\substack{(\varGamma_2'',\varGamma_2')\in\langle\varGamma_2\rangle\\(\varGamma_1'',\varGamma_1')\in\langle\varGamma_1\rangle}}\ \ \sum_{\substack{\varGamma''\in\varGamma_2''\plug{}\varGamma_1''\\\varGamma'\in\varGamma_2'\plug{}\varGamma_1'}}\varGamma''\otimes\varGamma'\ .\label{DM2}
\end{eqnarray}
A closer look at condition \texttt{(4)} and Eq.~(\ref{compatibility}) shows a one-to-one correspondence between terms in the sums on the right-hand sides of Eqs.~(\ref{DM1}) and (\ref{DM2}), verifying the validity of Eq.~(\ref{delta-morph}).

Verification of Eq.~(\ref{epsilon-morph}) rests upon the simple observation that composition of diagrams $\varGamma_2*\varGamma_1$ yields the void diagram only if both of them are void. Then, both sides are equal to 1 if $\varGamma_1=\varGamma_2=\text{\O}$ and 0 otherwise, which confirms Eq.~(\ref{epsilon-morph}).

\subsection*{\emph{(iii)} Hopf algebra}

A Hopf algebra structure consists of a bi-algebra $(\mathcal{G},+,*,\text{\O},\Delta,\varepsilon)$ equipped with an antipode $S:\mathcal{G}\longrightarrow\mathcal{G}$ which is an endomorphism satisfying the property
\begin{eqnarray}\label{antipode-def}
\mu\circ(Id\otimes S)\circ\Delta=\Xi=\mu\circ(S\otimes Id)\circ\Delta\ ,
\end{eqnarray}
where $\mu:\mathcal{G}\otimes\mathcal{G}\longrightarrow\mathcal{G}$ is the multiplication $\mu(\varGamma_2\otimes \varGamma_1)=\varGamma_2*\varGamma_1$, and $Id:\mathcal{G}\longrightarrow\mathcal{G}$ is the identity map on $\mathcal{G}$.  We have introduced the auxiliary linear mapping $\Xi:\mathcal{G}\longrightarrow\mathcal{G}$ merely to simplify the proof. This mapping  is defined by
 $\Xi=\eta \circ \,\varepsilon$ where the unit map $\eta:\mathbb{K}\longrightarrow \mathcal{G}$ satisfies $\eta(\alpha)=\alpha \text{\O}$. $\Xi$  is thus the projection on the subspace spanned by \O, \textit{i.e.}
\begin{eqnarray}\label{epsilon-proj}
\Xi(\varGamma)=\left\{
\begin{array}{l}
\varGamma \text{\ \ \ \ if\ \ }\varGamma=\alpha\,\text{\O}\ ,\; \;  \alpha \in \mathbb{K}\ , \\
0  \text{\ \ \ \, otherwise\ . }
\end{array}\right.
\end{eqnarray}

We now prove that $S$ given in Eq.~(\ref{antipode}) satisfies the condition of Eq.~(\ref{antipode-def}). We start by considering an auxiliary linear mapping $\Phi:\mathsf{End}(\mathcal{G})\longrightarrow \mathsf{End}(\mathcal{G})$ defined by
\begin{eqnarray}\label{Phi}
\Phi(f)=\mu\circ(Id\otimes f)\circ\Delta,\ \ \ \ \ \ f\in \mathsf{End}(\mathcal{G}).
\end{eqnarray}
Observe that under the assumption that $\Phi$ is invertible the first equality in Eq.~(\ref{antipode-def}) can be rephrased into the condition
\begin{eqnarray}\label{antipode-inverse-Phi}
S=\Phi^{-1}(\Xi)\ .
\end{eqnarray}
Now, our objective is to show that $\Phi$ is invertible and calculate its inverse explicitly.
By extracting the identity we get $\Phi=Id+\Phi^+$ and observe that such defined $\Phi^+$ can be written in the form
\begin{eqnarray}
\Phi^+(f)=\mu\circ(\bar{\Xi}\otimes f)\circ\Delta,\ \ \ \ \ \ f\in \mathsf{End}(\mathcal{G})\ ,
\end{eqnarray}
where $\bar{\Xi}=Id-\Xi$ is the complement of $\Xi$ projecting on the subspace spanned by $\varGamma\neq\text{\O}$, \textit{i.e.}
\begin{eqnarray}\label{epsilon-proj-complement}
\bar{\Xi}(\varGamma)=\left\{
\begin{array}{l}
0 \text{\ \ \ \ \ if\ \ }\varGamma=\alpha\,\text{\O}\ ,\; \;  \alpha \in \mathbb{K}\ , \\
\varGamma  \text{\ \ \ \ otherwise\ . }
\end{array}\right.
\end{eqnarray}
We claim that the mapping $\Phi$ is invertible with inverse given by \footnote{For a linear mapping $L=Id+L^+:V\longrightarrow V$ its inverse can be constructed as $L^{-1}=\sum_{n=0}^\infty (-L^+)^n$ provided the sum is well defined. Indeed, one readily checks that $L\circ L^{-1}=(Id+L^+)\circ\sum_{n=0}^\infty (-L^+)^n=\sum_{n=0}^\infty (-L^+)^n+\sum_{n=0}^\infty (-L^+)^{n+1}=Id$, and similarly $L^{-1}\circ L=Id$.}
\begin{eqnarray}\label{Phi-1}
\Phi^{-1}=\sum_{n=0}^\infty \ (-\Phi^+)^n\ .
\end{eqnarray}
In order to check that the above sum is well defined we analyze the sum term by term. It is not difficult to calculate powers of $\Phi^+$ explicitly
\begin{eqnarray}\label{Phi+f}
\!\!\!\!\!\left(\Phi^+\right)^n(f)(\varGamma)=\!\!\!\!\!
\sum_{\substack{
\varGamma\leadsto(\varGamma_n,...,\varGamma_1,\varGamma_0)\\
\varGamma_n,...,\varGamma_1\neq\text{\O}}}\!\!\!\!\!
\varGamma_n*...*\varGamma_1*f(\varGamma_0)\,.
\end{eqnarray}
We note that in the above formula products of multiple decompositions arise from repeated use of the property of Eq.~(\ref{delta-morph}); the exclusion of empty components in the decompositions (except the single one on the right hand side) comes from the definition of $\bar{\Xi}$ in Eq.~(\ref{epsilon-proj-complement}). The latter constraint together with condition \texttt{(5)} asserts that the number of non-vanishing terms in Eq.~(\ref{Phi-1}) is always finite proving that $\Phi^{-1}$ is well defined.
Finally, using Eqs.~(\ref{Phi-1}) and (\ref{Phi+f}) one explicitly calculates $S$ from Eq.~(\ref{antipode-inverse-Phi}), obtaining the formula of Eq.~(\ref{antipode}).

In conclusion, by construction the linear mapping $S$ of Eq.~(\ref{antipode}) satisfies the first equality in Eq.~(\ref{antipode-def}); the second equality can be checked analogously. Therefore we have proved $S$ to be an antipode thus making $\mathcal{G}$ into a Hopf algebra. We remark that, by a general theory of Hopf algebras \cite{SweedlerBook,AbeBook}, the property of Eq.~(\ref{antipode-def}) implies that $S$ is an anti-morphism and that it is unique. Moreover, if $\mathcal{G}$ is commutative or co-commutative $S$ is an involution, \textit{i.e.} $S\circ S=Id$.

\section{Properties of diagram decomposition}\label{PropDiagDecomp}
We verify that the decomposition of Definition~\ref{HWDiag-decomp} satisfies conditions \texttt{(0)} - \texttt{(5)} of Section~\ref{BasicNotions}.

Condition \texttt{(0)} follows directly from the construction, as we consider finite diagrams only.

\begin{figure}[h]
\begin{center}
\resizebox{0.4\textwidth}{!}{\includegraphics{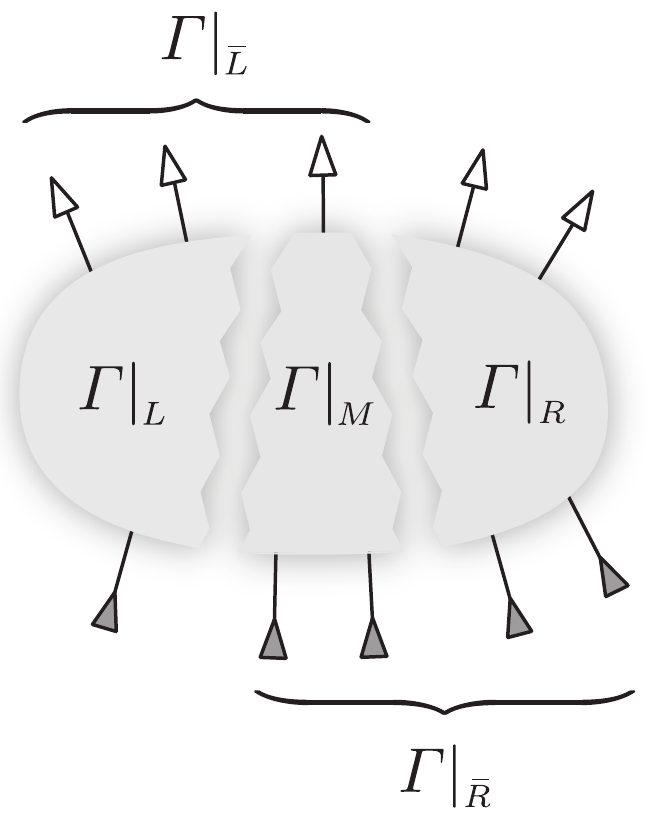}}
\caption{\label{Diagram-TripleDecomp} Triple decomposition of a Heisenberg--Weyl diagram used in the proof of condition \texttt{(1)}.}
\end{center}
\end{figure}

The proof of condition \texttt{(1)} consists of providing a one-to-one correspondence between schemes (\ref{TripleDecomp1}) and (\ref{TripleDecomp2}) decomposing a diagram $\varGamma$ into triples. Accordingly, one easily checks (see illustration Fig.~\ref{Diagram-TripleDecomp}) that each triple $(\left.\varGamma\right|_L,\left.\varGamma\right|_{M},\left.\varGamma\right|_{R})$ obtained by
\begin{eqnarray}
\varGamma\leadsto(\left.\varGamma\right|_L,\left.\varGamma\right|_{\bar{R}})\leadsto(\left.\varGamma\right|_L,\left.\varGamma\right|_{M},\left.\varGamma\right|_{R}) \,\label{D1}
\end{eqnarray}
where $\left.\varGamma\right|_{\bar{R}}\leadsto (\left.\varGamma\right|_{M},\left.\varGamma\right|_{R})$, also turns up as the decomposition
\begin{eqnarray}
\varGamma\leadsto(\left.\varGamma\right|_{\bar{L}},\left.\varGamma\right|_R)\leadsto(\left.\varGamma\right|_{L},\left.\varGamma\right|_{M},\left.\varGamma\right|_{R})\label{D2}\ ,
\end{eqnarray}
where $\left.\varGamma\right|_{\bar{L}}\leadsto (\left.\varGamma\right|_{L},\left.\varGamma\right|_{M})$, for the choice $\bar{L}=L+M$. Conversely, triples obtained by the scheme (\ref{D2}) coincide with the results of (\ref{D1}) for the choice $\bar{R}=M+R$. Therefore, the multisets of triple decompositions are equal and Eq.~(\ref{TripleDecomp}) holds.

Condition \texttt{(2)} is straightforward since the void graph \O\ is given by the empty set of lines, and hence the decompositions $\varGamma\leadsto(\varGamma,\text{\O})$ and $\varGamma\leadsto(\text{\O},\varGamma)$ are uniquely defined by the partitions $E_\varGamma+\text{\O}=E_\varGamma$ and $\text{\O}+E_\varGamma=E_\varGamma$ respectively.

The symmetry condition \texttt{(3)} results from swapping subsets $L\leftrightarrow R$ in the partition $L+R=E_\varGamma$ which readily yields Eq.~(\ref{symmetry}).

\begin{figure}[h]
\begin{center}
\resizebox{0.8\textwidth}{!}{\includegraphics{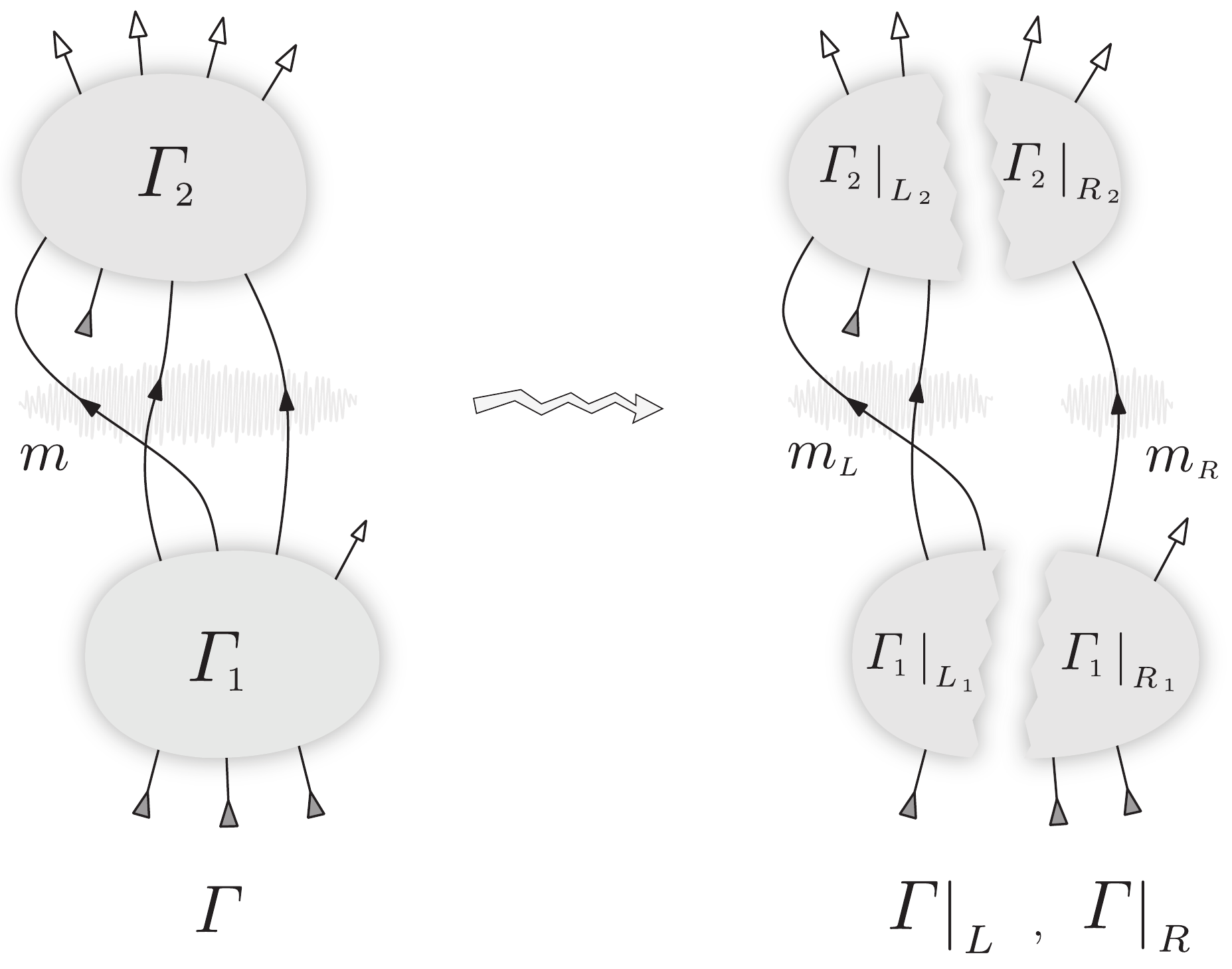}}
\caption{\label{Diagram-decomp-proof} Decompositions of a composite diagram $\varGamma=\varGamma_2\plug{m}\varGamma_1$ for some $m\in\varGamma_2\match{\ }\varGamma_1$ used in the proof of condition \texttt{(4)}.}
\end{center}
\end{figure}

In order to check property \texttt{(4)} we need to construct a one-to-one correspondence between elements of both sides of Eq.~(\ref{compatibility}). First, we observe that elements of the left-hand-side are decompositions of $\varGamma_2\plug{m}\varGamma_1$ for all $m\in\varGamma_2\match{\ }\varGamma_1$, \textit{i.e.}
\begin{eqnarray}\label{Comp1}
(\varGamma_2\plug{m}\left.\varGamma_1\right|_L,\varGamma_2\plug{m}\left.\varGamma_1\right|_R)
\end{eqnarray}
where $L+R=E_{\varGamma_2\plugsubs{m}\varGamma_1}$.
On the other hand, the right-hand-side consists of component-wise compositions of pairs $(\left.\varGamma_2\right|_{L_2},\left.\varGamma_2\right|_{R_2})\in\langle\varGamma_2\rangle$ and $(\left.\varGamma_1\right|_{L_1},\left.\varGamma_1\right|_{R_1})\in\langle\varGamma_1\rangle$ for $L_2+R_2=E_{\varGamma_2}$ and $L_1+R_1=E_{\varGamma_1}$, which written explicitly are of the form
\begin{eqnarray}\label{Comp2}
(\left.\varGamma_2\right|_{L_2}\plug{m_L}\left.\varGamma_1\right|_{L_1},\left.\varGamma_2\right|_{R_2}\plug{m_R}\left.\varGamma_1\right|_{R_1})
\end{eqnarray}
with $m_L\in\left.\varGamma_2\right|_{L_2}\match{\ }\left.\varGamma_1\right|_{L_1}$ and $m_R\in\left.\varGamma_2\right|_{R_2}\match{\ }\left.\varGamma_1\right|_{R_1}$.
We construct two mappings between elements of type (\ref{Comp1}) and (\ref{Comp2}) by the following assignments, see Fig.~\ref{Diagram-decomp-proof} for a schematic illustration. The first one is defined as:
\begin{eqnarray}
(m,L,R)\longrightarrow(L_1,R_1,L_2,R_2,m_L,m_R)\ ,\nonumber
\end{eqnarray}
where $L_i=E_{\varGamma_i}\cap L$, $R_i=E_{\varGamma_i}\cap R$ for $i=1,2$ and $m_L=m\cap L$, $m_R=m\cap R$.  The second one is given by:
\begin{eqnarray}
(L_1,R_1,L_2,R_2,m_L,m_R)\longrightarrow(m,L,R)\ ,\nonumber
\end{eqnarray}
with $m=m_L\cup m_R$ and $L=L_2\cup L_1$, $R=R_2\cup R_2$.
One checks that these mappings are inverses of each other and, moreover, the corresponding pairs of diagrams (\ref{Comp1}) and (\ref{Comp2}) are the same. This verifies that the multisets on the left- and right-hand sides of Eq.~(\ref{compatibility}) are equal and that condition \texttt{(4)} is satisfied.

Condition \texttt{(5)} is straightforward from the construction since the edges of a diagram $\varGamma$ can be nontrivially partitioned into at most  $|\varGamma|$ subsets (each consisting of  one edge only).

\pagebreak

\bibliography{Bibliography}
\bibliographystyle{elsarticle-num-names}

\end{document}